\documentclass[12pt]{article}

\usepackage[centertags]{amsmath}
\usepackage{amsmath}
\usepackage{amsfonts}
\usepackage{amssymb}
\usepackage{amsthm}
\usepackage{newlfont}
\usepackage{epsfig}
\usepackage{amscd}
\usepackage{pstricks}
\usepackage{epsfig}
\usepackage{graphicx}
\usepackage{color}
\usepackage{cite}

\newcommand{\NN}{{\mathbb N}}

\newcommand{\RR}{{\mathbb R}}

\newcommand{\beq}{\begin{equation}}
\newcommand{\eeq}{\end{equation}}
\newcommand{\ba}{\begin{array}}
\newcommand{\ea}{\end{array}}
\newcommand{\bea}{\begin{eqnarray}}
\newcommand{\eea}{\end{eqnarray}}
\newcommand{\eps}{{\epsilon}}

\usepackage[centertags]{amsmath}
\usepackage{amsfonts}
\usepackage{amssymb}
\usepackage{amsthm}
\usepackage{newlfont}
\usepackage{epsfig}
\usepackage{amscd}

\begin{document}

\begin{center}
{\large \sc \bf The Inverse Spectral Transform for the \\ Dunajski hierarchy and some of its reductions, I: \\
Cauchy problem and longtime behavior of solutions}

\vskip 20pt

{\large G. Yi $^{1,\S}$ and P. M. Santini$^{2,\S}$}

\vskip 20pt

{\it
$^1$ School of Mathematical Sciences, Huaqiao University, Quanzhou 362021, China \\
$^2$ Dipartimento di Fisica, Universit\`a di Roma "La Sapienza", and \\
Istituto Nazionale di Fisica Nucleare, Sezione di Roma 1 \\
Piazz.le Aldo Moro 2, I-00185 Roma, Italy}

\bigskip

$^{\S}$e-mail:  {\tt ge.yi@hqu.edu.cn, paolo.santini@roma1.infn.it}

\bigskip

{\today}

\end{center}

\begin{abstract}

In this paper we apply the formal Inverse Spectral Transform for integrable dispersionless PDEs arising from the commutation condition of pairs of one-parameter families of vector fields, recently developed by S. V. Manakov and one of the authors, to one distinguished class of equations, the so-called Dunajski hierarchy. We concentrate, for concreteness, i) on the system of PDEs characterizing a general anti-self-dual conformal structure in neutral signature, ii) on its first commuting flow, and iii) on some of their basic and novel reductions. We formally solve their Cauchy problem and we use it to construct the longtime behavior of solutions, showing, in particular, that unlike the case of soliton PDEs, different dispersionless PDEs belonging to the same hierarchy of commuting flows evolve in time in very different ways, exhibiting either a smooth dynamics or a gradient catastrophe at finite time.

\end{abstract}

\section{Introduction}

Waves propagating in weakly nonlinear and dispersive media are well described by integrable soliton equations, like the Korteweg - de Vries \cite{KdV} equation and its integrable $(2+1)$ dimensional generalization, the Kadomtsev - Petviashvili \cite{KP}  equation. The Inverse Spectral Transform (IST), introduced by Gardner, Green, Kruskal and Miura \cite{GGKM}, is the spectral method allowing one to solve the Cauchy problem for such PDEs, predicting that a localized disturbance evolves into a number of soliton pulses + radiation, and solitons arise as an exact balance between nonlinearity and dispersion \cite{ZMNP},\cite{AS},\cite{CD},\cite{AC}. Soliton PDEs arise in hierarchies of commuting flows, and equations of the same hierarchy share the same multisoliton solution and similar behavior. It is known that, apart from exceptional cases, soliton PDEs do not generalize naturally to more than $(2+1)$ dimensions; therefore, in the context of soliton equations, integrability is a property of low dimensional PDEs.

There is another important class of integrable PDEs, the so-called dispersionless PDEs (dPDEs), or PDEs of hydrodynamic type, including as distinguished multidimensional examples the dispersionless Kadomtsev-Petviashvili (dKP) equation \cite{Timman1962,Zobolotskaya1969}, describing weakly nonlinear and quasi one dimensional waves in Nature \cite{Timman1962,Zobolotskaya1969,MS_dKPn_11}, the heavenly equation, arising from the the vacuum Einstein equations and the conformal anti-self-duality condition for the signature metric in canonical Pleb\'{a}nski form \cite{Plebanski}, the Dunajski equation \cite{Dunaj}, an integrable generalization of the heavenly equation in which only the anti-self-duality condition in kept, the dispersionless 2D Toda (d2DT) equation \cite{FP,Zak,BF}, whose elliptic and hyperbolic versions are both relevant, describing, for instance, integrable ${\cal H}$-spaces (heavens) \cite{BF,GD}, integrable Einstein - Weyl geometries \cite{H}-\cite{J},\cite{Ward}, and playing a key role in the study of the ideal Hele-Shaw problem \cite{MWZ,WZ,KMZ,LBW,MAM}; the Pavlov equation \cite{Pavlov}, \cite{Ferapontov}, \cite{Duna}, arising in the study of integrable hydrodynamic chains, and the Manakov-Santini system \cite{MS_dKP_IST_06}, giving a local description of any Lorentzian Einstein-Weyl geometry \cite{DuFerK}, and including, as particular cases, the dKP and the Pavlov equations. dPDEs arise, more in general, in various problems of Mathematical Physics and are intensively studied in the recent literature (see, f.i., \cite{ZS,Zak,KG,Kri0,Kri1,Taka1,TT1,TT2,Zakharov,Kri2,TT3,Penrose1976,DuMa1,DuMa2,DMT,DT,K-MA-R,MA-S,G-M-MA,Pavlov,Duna,Ferapontov,NNS,BK,KM1,KM2,BK2,DuFerK})). Since they arise from the condition of commutation $[\hat L,\hat M]=0$ of pairs of one-parameter families of vector fields, implying the existence of common zero energy eigenfunctions:
\beq\label{Lax_pair}
[\hat L,\hat M ]=0~~\Rightarrow~~\hat L\psi=\hat M\psi=0,
\eeq
they can be in an arbitrary number of dimensions \cite{ZS}. In addition, due to the lack of dispersion, these multidimensional PDEs may or may not exhibit a gradient catastrophe at finite time and, as we shall see in this paper, even in the same hierarchy there exist equations evolving in a smooth way or into a gradient catastrophe at finite time. Their integrability gives a unique chance to study analytically such a mechanism and, also with this motivation, a novel IST for vector fields, significantly different from that of soliton PDEs \cite{ZMNP,AS,AC}, has been recently constructed \cite{MS_shiftedRH_05,MS_IST_heav_06,MS_dKP_IST_06,MS_PMNP13_14}, at a formal level, by Manakov and one of the authors of this paper (PMS), i) to solve their Cauchy problem \cite{MS_shiftedRH_05,MS_IST_heav_06,MS_dKP_IST_06,MS_Pavlov_IST_07,MS_d2DT_09}, ii) obtain the longtime behavior of solutions\cite{MS_dKP_breaking_08,MS_heav_pavlov_09,MS_d2DT_09}, iii) costruct distinguished classes of exact implicit solutions \cite{MS_dKP_breaking_08,MS_heav_pavlov_09,MS_d2DT_09,MS_solv_11}, see also \cite{BDM}, iv) establish if, due to the lack of dispersion, the nonlinearity of the PDE is ``strong enough'' to cause the gradient catastrophe of localized multidimensional disturbances, and v) study analytically the breaking mechanisms \cite{MS_dKP_breaking_08,MS_dKPn_11,MS_finite_t_12}. It is important to mention that the main difficulty to make the above IST for vector fields rigorous is associated with the proof of existence of analytic eigenfunction, motivated by the small field limit in the spectral parameter $\lambda$. A proof of existence, but only for Im $\lambda > c$, was found in \cite{GS} for the IST of the dKP equation; a complete proof exists so far only in the simplest case of the Pavlov equation, for which the whole IST has been recently made rigorous \cite{GSW}.

In this paper we concentrate on the Dunajski hierarchy, whose elegant structure was enveiled in \cite{BDM,Bogdanov2}. More precisely, in \S 2 we consider some basic members of the Dunajski hierarchy, corresponding to the same Lax operator $\hat L$: i) the system of PDEs \cite{BDM} characterizing a general anti-self-dual conformal structure in neutral signature \cite{DuFerK}, ii) its first commuting flow, and iii) some of their basic and novel reductions. In \S 3 we construct the formal IST for the Dunajski hierarchy corresponding to the same Lax operator $\hat L$, and we use it to solve the Cauchy problem for localized (in $x,y,z$) initial data. While the direct and inverse problems are the same for all the equations of this hierarchy, being associated with the same Lax operator $\hat L$, the $t$-evolution of different equations of such hierarchy, ruled by different $\hat M$ operators, is considerably different, as we shall see in this paper, leading either to a smooth dynamics or  to a gradient catastrophe of dKP type \cite{MS_dKP_breaking_08,MS_finite_t_12}. In \S 4 we construct the nonlinear Riemann-Hilbert (RH) dressing for the above equations, connecting also the RH data to the initial data of the dPDEs. In \S 5 we discuss the constraints on the spectral data corresponding to the reductions of the first commuting flows of the hierarchy. In \S 6 we use the NRH problem to construct the longtime behavior of the solutions. In \S 7 we make some concluding remarks and we discuss interesting open problems to be investigated in future works.

This paper is dedicated to the memory of S. V. Manakov.

\section{The first members of the Dunajski hierarchy and their basic reductions}

The Dunajski hierarchy is a basic example of hierarchy of integrable dPDEs, including the heavenly and the Manakov-Santini hierarchies as particular cases; see \cite{BDM} for details and for an elegant characterization of it.

\subsection{The first two commuting flows}

The first and basic member of such hierarchy is the following system of three PDEs \cite{BDM}
\beq\label{uvf_system1}
\ba{l}
u_{xt_1}-u_{yz}-(u_{x}-v_{y})u_{xy}+u_{y}u_{xx}-v_{x}u_{yy}=f_{y}, \\
v_{xt_1}-v_{yz}-(u_{x}-v_{y})v_{xy}+u_{y}v_{xx}-v_{x}v_{yy}=-f_{x}, \\
f_{xt_1}-f_{yz}-(u_{x}-v_{y})f_{xy}+u_{y}f_{xx}-v_{x}f_{yy}=0,
\ea
\eeq
equivalent to the commutation $[\hat{L},\hat{M_1}]=0$ of the vector fields:
\beq
\ba{l}
\hat{L}=\partial_{z}+\lambda\partial_{x}+\hat u_{x}-f_{x}\partial_{\lambda}, \\
\hat{M}_{1}=\partial_{t_1}+\lambda\partial_{y}+\hat u_{y}-f_{y}\partial_{\lambda},
\ea
\eeq
where
\beq
\hat u=u\partial_x+v\partial_y .
\eeq
It was recently shown \cite{DuFerK} that there exist local coordinates $(t_1,z,x,y)$ such that any anti-self-dual conformal structure in signature $(2,2)$ is locally represented by the metric
\beq
g=dt_1 dx+dz dy +u_y dt^2_1-(u_x+v_y)dt_1 dz+v_y dz^2 ,
\eeq
where $u$ and $v$ satisfy equations (\ref{uvf_system1}).

Keeping $\hat L$ fixed and varying $\hat M$, one obtains a hierarchy of commuting flows. The first commuting flow reads
\beq
\ba{l}\label{uvf_system2a}
u_{xt_2}=(\partial_z+{\hat u}_x)\alpha-u_y f_x-u_{xy}\beta-u_{xx}\alpha -\gamma, \\
v_{xt_2}=(\partial_z+{\hat u}_x)\beta -v_y f_x -v_{xy}\beta -v_{xx}\alpha, \\
f_{xt_2}=-(\partial_z+{\hat u}_x)\gamma -f_x f_y -f_{xy}\beta-f_{xx}\alpha,
\ea
\eeq
where the fields $\alpha,\beta,\gamma$ are defined in terms of $u,v,f$ through the equations
\beq
\ba{l}\label{uvf_system2b}
\alpha_x= -u_{yz}-u_x u_{xy}-v_x u_{yy}+v_y u_{xy}+u_y u_{xx}-f_y, \\
\beta_x=-v_{yz}-u_x v_{xy}-v_x v_{yy}+v_y v_{xy}+u_y v_{xx}+2 f_x, \\
\gamma_x=f_{yz}+{\hat u}_xf_y -{\hat u}_y f_x.
\ea
\eeq
Equations (\ref{uvf_system2a}),(\ref{uvf_system2b}) arise as the commutation condition $[\hat{L},\hat {M}_2]=0$ of ${\hat L}$ with the vector field
\beq
\ba{l}
{\hat M}_2=\partial_{t_2}+(u_y\lambda+\alpha)\partial_x+(\lambda^2+v_y\lambda+\beta)\partial_y+(-f_y\lambda+\gamma)\partial_{\lambda},
\ea
\eeq
whose coefficients increase by one their degree as polynomials of $\lambda$, with respect to ${\hat M}_1$.

\subsection{Some basic reductions}

The Dunajski hierarchy admits interesting reductions (see, f.i. \cite{BDM,Bogdanov2}). The reductions of equations (\ref{uvf_system1}) and (\ref{uvf_system2a}),(\ref{uvf_system2b}) we consider in this paper are of two types. Arising from the commutation conditions $[\hat L,\hat M_n]=0$ for one parameter families of vector fields, the first and natural reduction considered here is the divergence free condition for such vector fields (the condition $\nabla\cdot\vec u=0$ for the vector field $\hat X=\partial_t+\vec u\cdot\nabla$), implying the constant volume condition for the associated dynamical systems. The second basic set of reductions explored in this work, giving rise to non-autonomous systems, are associated with the condition that the so-called Orlov eigenfunctions of $(\hat L,\hat M_n)$ be polynomial in the spectral parameter $\lambda$ (see \S 5 for more details).\\

Beginning with the system (\ref{uvf_system1}), the divergenceless constraint for $\hat L$ and $\hat M_1$ :
\beq\label{volumeVF}
u_x+v_y=0,~~\Rightarrow~~u=\theta_y,~v=-\theta_x
\eeq
leads to the well known Dunajski equation \cite{Dunaj}
\beq\label{Dunaj}
\ba{l}
\theta_{xt_1} -\theta_{yz} + \theta_{xx} \theta_{yy} - \theta^{2}_{xy} = f, \\
f_{xt_1} - f_{yz} + \theta_{yy} f_{xx} + \theta_{xx} f_{yy} -2 \theta_{xy} f_{xy}=0,
\ea
\eeq
and to its vector fields Lax pair:
\beq
\ba{l}
\hat{L} = \partial_{z}+ \lambda \partial_{x} + \theta_{xy} \partial_{x}+ \theta_{xx} \partial_{y}
- f_{x} \partial_{\lambda}, \\
\hat{M}_{1} = \partial_{t} + \lambda \partial_{y} + \theta_{yy} \partial_{x} - \theta_{xy} \partial_{y} - f_{y} \partial_{\lambda},
\ea
\eeq
reducing to the second heavenly equation
\beq\label{heavenly}
\theta_{xt_1} -\theta_{yz} + \theta_{xx} \theta_{yy} - \theta^{2}_{xy} =0
\eeq
for $f=0$.

The second set of reductions of system (\ref{uvf_system1}) lead to non-autonomous and, to the best of our knowledge, novel integrable nonlinear PDEs. \\

The reduction
\beq\label{red1}
u+zf=0,
\eeq
leads to the system
\beq\label{red1_equ}
\ba{l}
v_{xt_1}-v_{yz}+(z f_{x}+v_{y})v_{xy}-z f_{y}v_{xx}-v_{x}v_{yy}=-f_{x}, \\
f_{xt_1}-f_{yz}+(z f_{x}+v_{y})f_{xy}-z f_{y}f_{xx}-v_{x}f_{yy}=0.
\ea
\eeq

The reduction
\beq\label{red2}
v+t_1 f=0,
\eeq
plays a similar role and does not give anything new, leading to the system
\beq\label{red2_equ}
\ba{l}
u_{xt_1}-u_{yz}-(t_1 f_{y}+u_{x})u_{xy}+t_{1}f_{x}u_{yy}+u_{y}u_{xx}=f_{y}, \\
f_{xt_1}-f_{yz}-(t_1 f_{y}+u_{x})f_{xy}+t_1 f_{x}f_{yy}+u_{y}f_{xx}=0,
\ea
\eeq
equivalent to (\ref{red1_equ}) by the change of variables $u\leftrightarrow v,~z\leftrightarrow t_1,~x\leftrightarrow y$.\\

The combination of the reductions (\ref{volumeVF}) and (\ref{red1}):
\beq\label{comb1}
u_x+v_y=u+zf=0~~\Rightarrow~~u=z\tilde\theta_y,~v=-z\tilde\theta_x,~f=-\tilde\theta_y,
\eeq
leads to the following scalar non authonomous generalization of the second heavenly equation:
\beq\label{equ_theta_tilde}
\tilde\theta_{xt_1}-\tilde\theta_{yz}+z\tilde\theta_{xx}\tilde\theta_{yy}-z\tilde\theta^2_{xy}=0,
\eeq
while the combination of the reductions (\ref{red1}) and (\ref{red2})
\beq\label{comb2}
u+zf=v+t_1 f=0,
\eeq
leads to the scalar dPDE
\beq\label{equ_f}
f_{xt}-f_{yz}+zf_x f_{xy}+t_1 f_x f_{yy}-z f_y f_{xx}-t_1 f_y f_{xy}=0.
\eeq

Concentrating now on equations (\ref{uvf_system2a}),(\ref{uvf_system2b}), the divergenceless constraint for the vector fields $\hat L,\hat M_2$, expressed now by the equations
\beq\label{volumeVF2}
u_x+v_y=\alpha_x+(\beta-f)_y=0 ~~\Rightarrow~~u=\theta_y,~v=-\theta_x,~\alpha=\rho_y,~\beta=f-\rho_x
\eeq
leads to the system
\beq \label{red6_equ}
\ba{l}
\Big(\theta_{xt_2}-(\partial_z+\theta_{xy}\partial_x-\theta_{xx}\partial_y)\rho+\theta_{xy}f\Big)_x+(\partial_z+\theta_{xy}\partial_x-\theta_{xx}\partial_y)f=0, \\
f_{xt_2}+f_x f_y+(\partial_z+\theta_{xy}\partial_x-\theta_{xx}\partial_y)\gamma +(f-\rho_x) f_{xy}+\rho_y f_{xx}=0, \\
\ea
\eeq
where
\beq \label{red6_abg}
\ba{l}
\rho_x+\theta_{yz}+\theta^2_{xy}-\theta_{xx}\theta_{yy}+f=0, \\
\gamma_x=f_{yz}+2\theta_{xy}f_{xy}-\theta_{xx}f_{yy}-\theta_{yy}f_{xx},
\ea
\eeq
reducing to the second member of the heavenly hierarchy
\beq\label{second_heavenly}
\ba{l}
\theta_{xt_2}=(\partial_z+\theta_{xy}\partial_x-\theta_{xx}\partial_y)\rho, \\
\rho_x+\theta_{yz}+\theta^2_{xy}-\theta_{xx}\theta_{yy}=0
\ea
\eeq
for $f=0$ and, consequently, $\gamma=0$.

The reduction (\ref{red1}) leads instead to the system
\beq \label{red7_equ}
\ba{l}
v_{x t_2}=(\partial_z-zf_x\partial_x+v_x\partial_y)\beta-f_x v_y-z\gamma v_{xx}-\beta v_{xy}, \\
f_{xt_2}+f_{x}f_y+(\partial_z-zf_x\partial_x+v_x\partial_y)\gamma +\beta f_{xy}+z\gamma f_{xx}=0,
\ea
\eeq
where $\alpha=z\gamma$ and
\beq \label{red7_abg}
\ba{l}
\beta_x=-(\partial_z-zf_x\partial_x+v_x\partial_y)v_y+(-z f_y\partial_x+v_y\partial_y)v_x+2f_x, \\
\gamma_x=(\partial_z-zf_x\partial_x+v_x\partial_y)f_y-(-z f_y\partial_x+v_y\partial_y)f_x, \\
\ea
\eeq
and the combination of the above two reductions leads to the equations
\beq \label{red8_equ}
\ba{l}
\tilde\theta_{xt_2}-\tilde\theta_y\tilde\theta_{xy}=\tilde\rho_z+z\tilde\theta_{xy}\rho_x-z\tilde\theta_{xx}\rho_y, \\
\tilde\rho_x+\tilde\theta_{yz}+z{\tilde\theta}^2_{xy}-z\tilde\theta_{xx}\tilde\theta_{yy}=0,
\ea
\eeq
where (\ref{comb1}) hold and $\tilde\rho=z^{-1}\rho,~\alpha=z\tilde\rho_y,~\beta=-\tilde\theta_y-z\tilde\rho_x,~\gamma=\tilde\rho_y$.

At last, the (less obvious) differential reduction
\beq\label{red9}
\ba{l}
(1-2t_2 f_{y})v_x=2t_2(f_z+u_x f_x),
\ea
\eeq
leads to the system
\begin{eqnarray} \label{red9_equ}
&&(1-2t_2 f_{y})(u_{xt_2 }+u_{y}f_{x}+u_{xy}\beta+u_{xx}\alpha+\gamma-\alpha_{z}-u_{x}\alpha_{x})=2t_2 (f_{z}+u_{x}f_{x})\alpha_{y}, \nonumber\\
&&(1-2t_2 f_{y})(f_{xt_2 }+f_{x}f_{y}+f_{xy}\beta+f_{xx}\alpha+\gamma_{z}+u_{x}\gamma_{x})=-2t_2 (f_{z}+u_{x}f_{x})\gamma_{y}, \nonumber\\
&&
\end{eqnarray}
where
\begin{eqnarray} \label{red9_abg}
&&(1-2t_2 f_{y})(\alpha_{x}+u_{yz}+u_{x}u_{xy}-u_{y}u_{xx}+f_{y})=2t_2 [(\gamma+u_{y}f_{x})u_{xy}-(f_{z}+u_{x}f_{x})u_{yy}],\nonumber\\
&&(1-2t_2 f_{y})\beta=2f+2t_2 (f_{t_2 }+\alpha f_{x}),\nonumber\\
&&(1-2t_2 f_{y})(\gamma_{x}-f_{yz}-u_{x}f_{xy}+u_{y}f_{xx})=2t_2 [(f_{z}+u_{x}f_{x})f_{yy}-(\gamma+u_{y}f_{x})f_{xy}].\nonumber\\
&&
\end{eqnarray}

The combination of reductions (\ref{red1}) and (\ref{red9}) leads to the dPDE
\begin{eqnarray} \label{red10_equ}
&&(1-2t_2 f_{y})(f_{xt_2 }+f_{x}f_{y}+f_{xy}\beta+\gamma_{z}+zf_{xx}\gamma-zf_{x}\gamma_{x})=2t_2 (z f^{2}_{x}-f_{z})\gamma_{y}, \nonumber\\
&&
\end{eqnarray}
where $\alpha=z\gamma$ and
\begin{eqnarray} \label{red10_abg}
&&(1-2t_2 f_{y})\beta=2f+2t_2 (f_{t_2 }+z\gamma f_{x}),\nonumber\\
&&(1-2t_2 f_{y})(\gamma_{x}-f_{yz}+zf_{x}f_{xy}-zf_{y}f_{xx})=2t_2 [(f_{z}-zf^{2}_{x})f_{yy}-(\gamma-zf_{x}f_{y})f_{xy}]. \nonumber\\
&&
\end{eqnarray}

At last, the combination of reductions (\ref{volumeVF2}) and (\ref{red9}) leads to the PDEs
\beq \label{red6_equ}
\ba{l}
2t_2[\theta_{xt_2}-(\partial_z+\theta_{xy}\partial_x-\theta_{xx}\partial_y)\rho+\theta_{xy}f]-\theta_x=0, \\
(1-2t_2 f_y)\theta_{xx}+2t_2(f_z+\theta_{xy}f_x)=0, \\
\rho_x+\theta_{yz}+\theta^2_{xy}-\theta_{xx}\theta_{yy}+f=0,
\ea
\eeq
where (\ref{volumeVF2}) holds, and $\gamma =(\theta_{xy}f)_y-(\theta_{yy}f)_x-\theta_{xy}/(2t_2)$.

The solution of the Cauchy problem of all the above reductions, in terms of a nonlinear RH problem, will be presented in \S 5.

\section{IST  for the Dunajski hierarchy}
In this section we apply the IST method introduced in \cite{MS_shiftedRH_05,MS_IST_heav_06,MS_dKP_IST_06} to construct the formal solution
of the Cauchy problem for the systems (\ref{uvf_system1}) and (\ref{uvf_system2a})-(\ref{uvf_system2b}) in $(3+1)$ dimensions and for their reductions, within the class of rapidly decreasing real potentials $u, v, f$:
\begin{eqnarray}
&&u,v,f\rightarrow 0,~~(x^2+y^2+z^2) \rightarrow + \infty, \nonumber\\
&&u,v,f\in \mathbb{R},~~(x,y,z) \in \mathbb{R}^{3},~~t_{j} > 0;
\end{eqnarray}
here $t_j,~j=1,2$ are interpreted as time variables and $x,y,z$ as space variables.

\subsection{Basic eigenfunctions}

A basic role in this IST theory is played by the real Jost eigenfunctions  $\vec{\varphi}_{\pm}(x,y,z;\lambda) \in \mathbb{R}^{3}$ for $\lambda\in\RR$, the solutions of $\hat {L} \vec{\varphi}_{\pm}=\vec{0}$ defined by the asymptotic:
\begin{eqnarray} \label{jost}
\vec{\varphi}_{\pm}(x,y,z;\lambda)=\left(\begin{array}{c}
                                     \varphi_{\pm0}(x,y,z;\lambda) \\
                                    \varphi_{\pm1}(x,y,z;\lambda)\\
                                    \varphi_{\pm2}(x,y,z;\lambda)
                                   \end{array}\right)
 \rightarrow \vec{\xi} \equiv \left(\begin{array}{c}
                                                           \lambda \\
                                                           \xi \equiv x-\lambda z \\
                                                           y
                                                         \end{array}\right), z \rightarrow \pm \infty.
\end{eqnarray}
These Jost eigenfunctions are intimately connected to the dynamical system (in the time variable $z$):
\begin{eqnarray} \label{dynamic}
\frac{d \vec{x}}{dz}=\left(\begin{array}{c}
                                                           -f_{x}(x,y,z) \\
                                                           \lambda+u_{x}(x,y,z) \\
                                                           v_{x}(x,y,z)
                                                         \end{array}\right)
,
\end{eqnarray}
for the unknown  $\vec{x}(z)=(\lambda(z),x(z),y(z))^{T}$ in the following way.

Assuming that the potentials $u,v,f$ be smooth and sufficiently localized functions of $x,y,z$, it follows from ODE theory that the solution
\beq
\vec x(z)=\left(
\ba{c}
\Lambda(z;x_0,y_0,z_0,\lambda_0) \\
X(z;x_0,y_0,z_0,\lambda_0) \\
Y(z;x_0,y_0,z_0,\lambda_0)
\ea
\right)
\eeq
of (\ref{dynamic}), satisfying the initial condition $\left(\lambda(z_{0}),x(z_{0}),y(z_{0})\right)^{T}=\left(\lambda_{0},x_{0},y_{0}\right)^{T}$, exists unique, and it is globally defined for real $z$, with the asymptotic states $\lambda_{\pm},x_{\pm},y_{\pm}$:
\begin{eqnarray} \label{asympt}
\lambda &\sim& \lambda_{\pm}(x_{0},y_{0},z_{0},\lambda_{0}),~~z \to \pm \infty, \nonumber\\
x &\sim& z \lambda_{\pm}(x_{0},y_{0},z_{0},\lambda_{0})+x_{\pm}(x_{0},y_{0},z_{0},\lambda_{0}),~z \to \pm \infty,\nonumber\\
y &\sim& y_{\pm}(x_{0},y_{0},z_{0},\lambda_{0}),~~z \to \pm \infty.
\end{eqnarray}
These asymptotic states $\lambda_{\pm}(x_{0},y_{0},z_{0},\lambda_{0}),x_{\pm}(x_{0},y_{0},z_{0},\lambda_{0}),y_{\pm}(x_{0},y_{0},z_{0},\lambda_{0})$ are constants of motion for the dynamical system (\ref{dynamic}) when the point $(x_{0},y_{0},\\ z_{0},\lambda_{0})$ moves along the trajectories; therefore they are solutions of the vector field equation
\beq
\hat {L}\left(\begin{array}{c}
                                     \lambda_{\pm}(x,y,z,\lambda) \\
                                    x_{\pm}(x,y,z,\lambda) \\
                                    y_{\pm}(x,y,z,\lambda)
                                   \end{array}\right)=\vec 0
\eeq
and, due to (\ref{jost}) and (\ref{asympt}), they coincide with the real Jost eigenfunctions $\vec{\varphi}_{\pm}(x,y,z;\lambda) $:
\beq
\ba{ccc}
\varphi_{\pm 0}(x,y,z;\lambda)&=&\lambda_{\pm}(x,y,z,\lambda),\\
\varphi_{\pm 1}(x,y,z;\lambda)&=&x_{\pm}(x,y,z,\lambda),\\
\varphi_{\pm 2}(x,y,z;\lambda)&=&y_{\pm}(x,y,z,\lambda).
\ea
\eeq
A crucial role in the IST for the vector field $\hat {L}$ is also played by the analytic eigenfunctions $\vec{\psi}^{\pm}(x,y,z;\lambda)$, the solutions of  $\hat {L} \vec{\psi}^{\pm}=0$ analytic, respectively, in the upper and lower halves of complex
$\lambda$ plane, satisfying the asymptotics
\begin{eqnarray}
\vec{\psi}^{\pm}(x,y,z;\lambda) \rightarrow \vec{\xi},~~~~~(x^{2}+y^{2}+z^{2}) \rightarrow +\infty.
\end{eqnarray}
They can be characterized by the following integral equations
\begin{eqnarray}
&&\vec{\psi}^{\pm}(x,y,z;\lambda)+\int_{\mathbb{R}^{3}}dx'dy'dz'G^{\pm}(x-x',y-y',z-z';\lambda)\dot \nonumber\\
&&[u_{x'}(x',y',z')\partial_{x'}+v_{x'}(x',y',z')\partial_{y'}-f_{x'}(x',y',z')\partial_{\lambda}] \vec{\psi}^{\pm}(x',y',z';\lambda) \nonumber\\
&&=\vec{\xi},
\end{eqnarray}
for the analytic Green's functions of the undressed operator $(\partial_z +\lambda\partial_x)$:
\begin{eqnarray} \label{analyticGreen}
G^{\pm}(x,y,z;\lambda)=\pm \frac{\delta(y)}{2 \pi \textbf{i}[x-(\lambda+\textbf{i}\varepsilon) z]}.
\end{eqnarray}
Consequently, for $|\lambda| \gg 1$:
\begin{eqnarray}\label{psi_u}
\vec{\psi}^{\pm}(x,y,z;\lambda)=\vec{\xi}+\frac{\vec{Q}(x,y,z)}{\lambda}+O(\lambda^{-2}),
\end{eqnarray}
where
\begin{eqnarray}\label{def_Q}
\vec{Q}(x,y,z)=\left(\begin{array}{c}
                 f(x,y,z) \\
                 -u(x,y,z)-z f(x,y,z) \\
                 -v(x,y,z)
               \end{array}\right).
\end{eqnarray}
We observe that the analytic Green's functions (\ref{analyticGreen}) exhibit the following asymptotics for $z \rightarrow \pm \infty$
\begin{subequations}
\begin{eqnarray}
G^{\pm}(x-x',y-y',z-z';\lambda) &\rightarrow& \pm \frac{\delta(y)}{2 \pi \textbf{i}[\xi-\xi' \mp \textbf{i}\varepsilon]},~~z \rightarrow +\infty,
\end{eqnarray}
\begin{eqnarray}
G^{\pm}(x-x',y-y',z-z';\lambda) &\rightarrow& \pm \frac{\delta(y)}{2 \pi \textbf{i}[\xi-\xi' \pm \textbf{i}\varepsilon]},~~z \rightarrow -\infty,
\end{eqnarray}
\end{subequations}
where $\xi=x-\lambda z,~\xi'=x'-\lambda z'$, entailing that the $z \rightarrow +\infty$ asymptotics of $\vec{\psi}^{\pm}$ are analytic, respectively, in the lower and upper halves of the complex $\xi$-plane, while the $z \rightarrow -\infty$ asymptotics of $\vec{\psi}^{\pm}$ are analytic, respectively, in the upper and lower halves of the complex $\xi$-plane.

\subsection{Scattering and Spectral data}

The $z$ (time) scattering problem for the ODE system (\ref{dynamic}) allows one to construct the scattering data $\vec\sigma(\vec\xi)$,
defined by the $z\to +\infty$ limit of $\vec{\varphi}_{-}$:
\begin{eqnarray} \label{scattering}
\lim_{z \to +\infty} \vec{\varphi}_{-}(x,y,z;\lambda)=\vec S(\vec{\xi})=\vec{\xi}+\vec{\sigma}(\vec{\xi}).
\end{eqnarray}
The first part of the direct problem is the mapping from the real potentials $f,u$ and $v$ , functions of the three real variables $(x,y,z)$, to the real scattering vector $\vec{\sigma}$ defined in (\ref{scattering}), function of the real variables $\vec{\xi}=(\lambda,\xi,y)$ .

Together with the above scattering data, one defines also spectral data in the following way.
The ring property of the space of eigenfunctions allows one to express the analytic eigenfunctions in
terms of the Jost eigenfunctions, used as a basis for such a space, if $\lambda\in\RR$:
\beq\label{relation1}
\ba{l}
\vec{\psi}^{+}(x,y,z;\lambda)=\vec{\cal K}^+_-(\vec{\varphi}_{-}(x,y,z;\lambda))=\vec{\cal K}^-_+(\vec{\varphi}_{+}(x,y,z;\lambda)),
\ea
\eeq
\beq\label{relation2}
\ba{l}
\vec{\psi}^{-}(x,y,z;\lambda)=\vec{\cal K}^-_-(\vec{\varphi}_{-}(x,y,z;\lambda))=\vec{\cal K}^+_+(\vec{\varphi}_{+}(x,y,z;\lambda)),
\ea
\eeq
where
\beq
\vec{\cal K}^a_b(\vec\xi):=\vec\xi +\vec{\chi}^a_b(\vec\xi),~~~a,b=\pm
\eeq
The $z \rightarrow -\infty$ and $z \rightarrow +\infty$ limits of respectively the first and second equalities
in (\ref{relation1}) and (\ref{relation2}) imply
\beq
\lim_{z \to -\infty} \vec{\psi}^{\pm}=\vec{\xi}+\vec{\chi}^{\pm}_{-}(\vec{\xi}), \ \
\lim_{z \to +\infty} \vec{\psi}^{\pm}=\vec{\xi}+\vec{\chi}^{\mp}_{+}(\vec{\xi}),
\eeq
Therefore the analyticity properties of $\vec{\psi}^{\pm}$ established above imply that $\vec{\chi}_{-}^{+}(\vec{\xi})$ and $\vec{\chi}_{+}^{+}(\vec{\xi})$ are analytic in $Im ~\xi >0$ decaying at $\xi \rightarrow \infty$ like $O(\xi^{-1})$,  while $\vec{\chi}_{-}^{-}(\vec{\xi})$ and $\vec{\chi}_{+}^{-}(\vec{\xi})$ are analytic  in $Im ~\xi <0$ decaying at $\xi \rightarrow \infty$ like $O(\xi^{-1})$. In addition, taking the $z \rightarrow +\infty$ limit of the second of (\ref{relation1}), one obtains the following equation:
 \begin{eqnarray} \label{RH_problem_shift}
 \vec{\sigma}(\vec{\xi})+\vec{\chi}_{-}^{+}\left(\vec{\xi}+\vec{\sigma}(\vec{\xi})\right)-\vec{\chi}_{+}^{-}(\vec{\xi})=\vec{0},~~~\lambda\in\RR ,
 \end{eqnarray}
that must be viewed as three ``linear scalar Riemann-Hilbert (RH) problems in the variable $\xi$, with the given shift $\vec{\sigma}(\vec{\xi})$'' for the unknowns $\vec{\chi}_{-}^{+}$ and $\vec{\chi}_{+}^{-}$ (see, f. i., \cite{Gakhov} for the associated theory). Such a RH problems with a shift are equivalent to the following three linear Fredholm equations \cite{Gakhov}
\begin{eqnarray} \label{spectral}
 {{\chi}^{-}_{+}}_j(\vec{\xi})-\frac{1}{2\pi i}\int_{\mathbb{R}}K(\vec{\xi},\vec{\xi}'){{\chi}^{-}_{+}}_j(\vec{\xi}')d\xi' +M_j(\vec{\xi})=0,~~~j=0,1,2,~~~\lambda\in\RR,
\end{eqnarray}
where
\beq
\ba{l}
K(\vec{\xi},\vec{\xi}')=\frac{\partial S(\vec{\xi}')/\partial \xi'}{S(\vec{\xi}')-S(\vec{\xi})}-\frac{1}{\xi'-\xi}, \nonumber\\
M_j(\vec{\xi})=-\frac{1}{2}{\sigma}_j(\vec{\xi})+\frac{1}{2\pi i}P.V.\int_{\mathbb{R}}\frac{\partial S(\vec{\xi}')/\partial \xi'}{S(\vec{\xi}')-S(\vec{\xi})}{\sigma}_j(\vec{\xi}')d\xi',~~~j=0,1,2, \\
S(\vec{\xi})=\xi+\sigma_1(\vec{\xi}),~~S(\vec{\xi}')=\xi'+\sigma_1(\vec{\xi}'),~~\\
\vec{\xi}=(\lambda,\xi,y),~~~\vec{\xi}'=(\lambda,\xi',y).
\ea
\eeq
We remark that, if $u,v,f\in\RR$, then the vector fields are real for $\lambda\in\RR$, and
\beq\label{real_data}
\ba{l}
\vec\varphi_{\pm}\in\RR^3, ~~~\overline{\vec\psi^{-}}=\vec\psi^{+},~~~ \lambda\in\RR, \\
\vec\sigma\in\RR^3, ~~~\overline{\vec{\cal K}^{-}_a}=\vec{\cal K}^{+}_a, ~~~\overline{\vec{\chi}^{-}_a}=\vec{\chi}^{+}_a .
\ea
\eeq

Summarizing, the direct problem consists of the following steps: i) given the initial data $(u(x,y,z,0),v(x,y,z,0),f(x,y,z,0))$, one
constructs the scattering data $\vec\sigma(\vec\xi,0)$ from the solution of the ODE system (\ref{dynamic}); ii) known $\vec\sigma(\vec\xi,0)$, one solves the RH problem with a shift (\ref{RH_problem_shift}) constructing the spectral data $\vec{\chi}^{\pm}_{-}$ and
$\vec{\chi}^{\pm}_{+}$.

\subsection{Inverse problems}

\noindent
\textbf{First inversion}. Known the spectral datum $\vec{\chi}^+_{-}(\vec\xi)$, an inverse problem can be constructed from the first of equations (\ref{relation1}), observing that
\beq\label{projection}
\hat{P}^-_{\lambda}(\vec\varphi_- +\vec\xi +\vec{\chi}^+_{-}(\vec\varphi_- ))=\vec 0,
\eeq
where
\begin{eqnarray}
\hat{P}^{\pm}_{\lambda}g(\lambda) \equiv \pm \frac{1}{2\pi\mathbf{i}} \int_{\mathbb{R}} \frac{g(\lambda')}{\lambda'-(\lambda \pm \mathbf{i}\varepsilon)}d \lambda'
\end{eqnarray}
are the $(\pm)$ analyticity projectors to the upper and lower halves of the complex $\lambda$ plane.
Since $\hat{P}^{\pm}_{\lambda}f({\lambda})=\pm\frac{\mathbf{i}}{2}H_{\lambda}f(\lambda)+\frac{1}{2} f(\lambda)$, equation (\ref{projection}) is equivalent to the inverse problem formula
\begin{eqnarray}\label{inverse1}
 \vec{\varphi}_{-}(x,y,z;\lambda)+H_{\lambda}\Im{\vec{\chi}}_{-}^{+}(\vec{\varphi}_{-}(x,y,z;\lambda))
 +\Re{\vec{\chi}}_{-}^{+}(\vec{\varphi}_{-}(x,y,z;\lambda'))=\vec{\xi},
 \end{eqnarray}
  where ~$\Re{\vec{\chi}}_{-}^{+}$~ and ~$\Im{\vec{\chi}}_{-}^{+}$~ are the real and imaginary parts of $\vec{\chi}_{-}^{+}$, i.e., $\vec{\chi}_{-}^{+}=\Re{\vec{\chi}}_{-}^{+}+\textbf{i} \Im{\vec{\chi}}_{-}^{+}$,
and $H_{\lambda}g(\lambda)$ is the Hilbert transform operator
\begin{eqnarray}
H_{\lambda}g(\lambda) \equiv \frac{1}{\pi}P.V. \int_{\mathbb{R}} \frac{g(\lambda')}{\lambda-\lambda'}d \lambda'.
\end{eqnarray}
Equation (\ref{inverse1}) must be viewed as a vector nonlinear integral equation for the unknown Jost eigenfunctions $\vec\varphi_-$.

Once $\vec{\varphi}_{-}$ is reconstructed from $\vec{\chi}_{-}^{+}$ solving the nonlinear integral equation (\ref{inverse1}), $f,u,v$ are finally reconstructed from (see \cite{GSW})
\beq\label{reconstruction1}
\ba{l}
f=-\frac{1}{\pi}\int_{\RR}\Im {{\chi}^+_{-}}_1(\vec\varphi_-(x,y,z,\lambda))d\lambda , \\
v=\frac{1}{\pi}\int_{\RR}\Im {{\chi}^+_{-}}_3(\vec\varphi_-(x,y,z,\lambda))d\lambda , \\
u=-zf+\frac{1}{\pi}\int_{\RR}\Im {{\chi}^+_{-}}_2(\vec\varphi_-(x,y,z,\lambda))d\lambda.
\ea
\eeq
To obtain the reconstruction formulae (\ref{reconstruction1}), one subtracts equations (\ref{relation1}) and (\ref{relation2}), and then uses the reality constraints (\ref{real_data}):
\beq
\vec\psi^+-\vec\psi^-=\vec{\chi}^+_{-}(\vec\varphi_-)-\vec{\chi}^-_{-}(\vec\varphi_-)=2 i\Im \vec{\chi}^+_{-}(\vec\varphi_-),
\eeq
together with the analyticity properties of $\vec\psi^{\pm}$, to get
\beq
\vec\psi^{\pm}=\vec\xi+\frac{1}{\pi}\int_{\RR}\frac{\Im\vec{\chi}^+_{-}(\vec\varphi_-(x,y,z,\lambda')) }{\lambda'-(\lambda\pm i \eps)}d\lambda'.
\eeq
Equations (\ref{reconstruction1}) follow then from (\ref{def_Q}).

\bigskip

\noindent
\textbf{Second inversion: the nonlinear Riemann-Hilbert problem}. An alternative inversion is based on the nonlinear Riemann-Hilbert (NRH) inverse problem on the real $\lambda$ axis:
\begin{eqnarray} \label{RH1}
\vec{\psi}^{+}(\lambda)&=&\tilde{\vec{\mathcal{R}}}\left(\vec{\psi}^{-}(\lambda)\right),~~\lambda\in\mathbb{R}
\end{eqnarray}
or, in component form:
\begin{eqnarray} \label{RH2}
\left(\begin{array}{c}
                           \psi_{0}^{+}(\lambda) \\
                           \psi_{1}^{+}(\lambda) \\
                           \psi_{2}^{+}(\lambda)
                          \end{array}\right)
= \left(\begin{array}{c}
   \tilde{\mathcal{R}}_{0}\left(\psi_{0}^{-}(\lambda),\psi_{1}^{-}(\lambda),\psi_{2}^{-}(\lambda)\right) \\
   \tilde{\mathcal{R}}_{1}\left(\psi_{0}^{-}(\lambda),\psi_{1}^{-}(\lambda),\psi_{2}^{-}(\lambda)\right) \\
   \tilde{\mathcal{R}}_{2}\left(\psi_{0}^{-}(\lambda),\psi_{1}^{-}(\lambda),\psi_{2}^{-}(\lambda)\right)
 \end{array}\right),~~\lambda \in \mathbb{R},
\end{eqnarray}
where the RH data $\tilde{\vec{\mathcal{R}}}(\vec{\zeta})$ are constructed from the spectral data $\vec{\chi}^{a}_{b}(\vec\xi)$ via algebraic manipulations, eliminating $\vec\varphi_-$ from the first of equations (\ref{relation1}) and (\ref{relation2}) The solutions $\vec{\psi}^{\pm}(\lambda)=\left(\psi_{0}^{\pm}(\lambda),\psi_{1}^{\pm}(\lambda),\psi_{2}^{\pm}(\lambda)\right)^{T}\in \mathbb{C}^{3}$ are analytic respectively in the upper and lower halves of the complex $\lambda$ plane, with the following
normalization in the neighborhood of $\lambda=\infty$:
\begin{eqnarray}\label{normalization1}
\psi_{0}^{\pm}(\lambda)=\lambda+O(\lambda^{-1}),~~\psi_{1}^{\pm}(\lambda)=-z\lambda +x+O(\lambda^{-1}),\nonumber\\
\psi_{2}^{\pm}(\lambda)=y+O(\lambda^{-1}),~~|\lambda |\gg 1,
\end{eqnarray}
and the potentials $u,v,f$, are reconstructed from (\ref{psi_u}):
\begin{eqnarray} \label{normalization1}
f&=&\displaystyle\lim_{\lambda \to \infty}{\lambda\left(\psi_{0}^{\pm}(\lambda)-\lambda\right)}, \nonumber\\
u&=&-zf-\displaystyle\lim_{\lambda \to \infty}{\lambda\left(\psi_{1}^{\pm}(\lambda)-x+z\lambda\right)},\nonumber \\
v&=&\displaystyle\lim_{\lambda \to \infty}{\lambda\left(\psi_{2}^{\pm}(\lambda)-y\right)}.
\end{eqnarray}

\subsection{Evolution of the scattering data}
In all formulas of the direct and inverse problems we have deliberately omitted the time variable, appearing just as a parameter. Now it is the time to introduce it, and, as we shall see, i) the evolution of the data is given by explicit formulae (and this is the main justification for the introduction of the IST); ii) such formulae imply a substantially different evolution for different members of the hierarchy; iii) the $t_1$ and $t_2$ flows commute on the level of the data, and this is the simplest way to prove that the corresponding flows (\ref{uvf_system1}) and (\ref{uvf_system2a}),(\ref{uvf_system2b}) commute as well.

It turns out that, as $f,u,v$ evolve in time according to (\ref{uvf_system1}), and, respectively, (\ref{uvf_system2a}),(\ref{uvf_system2b}), the time dependence of the three components of the data $\vec{S}$, $\vec{\cal K}^a_{b}$ and $\tilde{\vec{\cal R}}$ are defined by the same PDEs:
\beq\label{data_evol_1a}
\ba{l}
(\partial_{t_n}+\lambda^{n} \partial_{\zeta_2})D_j=0,~~j=0,1, \\
(\partial_{t_n}+\lambda^{n} \partial_{\zeta_2})D_2=D^n_0,
\ea
\eeq
whose explicit solutions read
\beq\label{data_evol_1b}
\ba{l}
D_j(\zeta_0,\zeta_1,\zeta_2,t_{n})=D_j(\zeta_0,\zeta_1,\zeta_2-t_n\zeta^n_0,0),~~j=0,1, \\
D_2(\zeta_0,\zeta_1,\zeta_2,t_{n})=D_2(\zeta_0,\zeta_1,\zeta_2-t_n\zeta^n_0,0)+t_n (D_0(\zeta_0,\zeta_1,\zeta_2-t_{n}\zeta^{n}_0,0))^n,
\ea
\eeq
where $n=1$ and, respectively, $n=2$. For an arbitrary $n\in\NN^+$ in (\ref{data_evol_1a}) and (\ref{data_evol_1b}), one generates the time dependence of the spectral data of a sequence of commuting flows of the Dunajski hierarchy.

Equations (\ref{data_evol_1a}) follow from the observation that one constructs, from the Jost $\vec\varphi_-$ and analytic $\vec\psi^{\pm}$ eigenfunctions of $\hat L$, the common Jost $\vec\phi_-$ and analytic $\vec\Psi^{\pm}$ eigenfunctions of both $\hat L$ and $\hat M_n$ operators via the formulae:
\begin{eqnarray}
\phi_{1}(x,y,z,t_n;\lambda) &\equiv& \varphi_{-0}(x,y,z,t_n;\lambda), \nonumber\\
\phi_{2}(x,y,z,t_n;\lambda) &\equiv& \varphi_{-1}(x,y,z,t_n;\lambda), \nonumber\\
\phi_{3}(x,y,z,t_n;\lambda) &\equiv& \varphi_{-2}(x,y,z,t_n;\lambda)-t_n (\varphi_{-0}(x,y,z,t_n;\lambda))^n
\end{eqnarray}
\begin{eqnarray}\label{connect_eigefunctions_1}
\Psi^{\pm}_0(x,y,z,t_n;\lambda) &\equiv& \psi^{\pm}_0(x,y,z,t_n;\lambda), \nonumber\\
\Psi^{\pm}_1(x,y,z,t_n;\lambda) &\equiv& \psi^{\pm}_1(x,y,z,t_n;\lambda), \nonumber\\
\Psi^{\pm}_2(x,y,z,t_n;\lambda) &\equiv& \psi^{\pm}_2(x,y,z,t_n;\lambda)-t_n (\psi^{\pm}_0(x,y,z,t_n;\lambda))^n,
\end{eqnarray}
using arguments developed, f.i., in \cite{MS_dKP_IST_06}.

It is straightforward to verify that the flows (\ref{data_evol_1a}) for different $n$ commute.

\section{The nonlinear RH dressing}

Since the RH data $\tilde{\vec{\cal R}}(\vec\zeta,t_n),~n=1,2$ satisfy equations (\ref{data_evol_1a}),(\ref{data_evol_1b}) and the analytic eigenfunctions $\vec\psi^{\pm}$ of $\hat L$ are connected to the analytic eigenfunctions $\vec\Psi^{\pm}$ of $\hat L,\hat M_n,~n=1,2$ through the equations (\ref{connect_eigefunctions_1}), it follows that the NRH formulation of the inverse problem, corresponding to the analytic eigenfunctions of both operators $\hat L,\hat M_n,~n=1,2$ reads as follows.
\vskip 10pt
\noindent
Consider the NRH problem
\begin{eqnarray} \label{RH1}
\vec{\Psi}^{+}(\lambda)&=&\vec{\mathcal{R}}\left(\vec{\Psi}^{-}(\lambda)\right),~~\lambda\in\mathbb{R}
\end{eqnarray}
or, in component form:
\begin{eqnarray} \label{RH2}
\left(\begin{array}{c}
                           \Psi_{0}^{+}(\lambda) \\
                           \Psi_{1}^{+}(\lambda) \\
                           \Psi_{2}^{+}(\lambda)
                          \end{array}\right)
= \left(\begin{array}{c}
   \mathcal{R}_{0}\left(\Psi_{0}^{-}(\lambda),\Psi_{1}^{-}(\lambda),\Psi_{2}^{-}(\lambda)\right) \\
   \mathcal{R}_{1}\left(\Psi_{0}^{-}(\lambda),\Psi_{1}^{-}(\lambda),\Psi_{2}^{-}(\lambda)\right) \\
   \mathcal{R}_{2}\left(\Psi_{0}^{-}(\lambda),\Psi_{1}^{-}(\lambda),\Psi_{2}^{-}(\lambda)\right)
 \end{array}\right),~~\lambda \in \mathbb{R},
\end{eqnarray}
where $\mathcal{R}_{j}(\vec\zeta)=\tilde{\mathcal{R}}_{j}(\vec\zeta,0)$, the solutions $\vec{\Psi}^{\pm}(\lambda)=\left(\Psi_{0}^{\pm}(\lambda),\Psi_{1}^{\pm}(\lambda),\Psi_{2}^{\pm}(\lambda)\right)^{T}
\in \mathbb{C}^{3}$ are analytic rispectively in the upper and lower halves of the complex $\lambda$ plane, with the following
normalizations in a neighborough of $\lambda=\infty$:
\beq\label{large_lambda}
\Psi^{\pm}_j(\lambda)=\nu_j(\lambda)+O(\lambda^{-1}),~~j=0,1,2,
\eeq
where, respectively,
\beq\label{norm1}
\nu_0(\lambda)=\lambda,~~\nu_1(\lambda)=x-\lambda z,~~\nu_2(\lambda)=y-\lambda t_1,
\eeq
and
\beq\label{norm2}
\nu_0(\lambda)=\lambda,~~\nu_1(\lambda)=x-\lambda z,~~\nu_2(\lambda)=-\lambda^2 t_2 +y-2f t_2.
\eeq
Then $\vec{\Psi}^{\pm}(\lambda)$ are vector eigenfunctions, respectively, of the vector fields $\hat L,\hat M_1$: $\hat L\vec{\Psi}^{\pm}=\hat M_1\vec{\Psi}^{\pm}=\vec 0$,  and of the vector fields $\hat L,\hat M_2$: $\hat L\vec{\Psi}^{\pm}=\hat M_2\vec{\Psi}^{\pm}=\vec 0$; furthemore, the fields $f,u,v$, reconstructed by, respectively, the following formulae:
\beq\label{reconstruct1_fuv}
\ba{l}
f=\displaystyle\lim_{\lambda \to \infty}{\lambda\left(\Psi_{0}^{\pm}(\lambda)-\lambda\right)}, \\
u=-zf-\displaystyle\lim_{\lambda \to \infty}{\lambda\left(\Psi_{1}^{\pm}(\lambda)-x+z\lambda\right)}, \\
v=-t_1f-\displaystyle\lim_{\lambda \to \infty}{\lambda\left(\Psi_{2}^{\pm}(\lambda)-y+t_1\lambda\right)},
\ea
\eeq
and
\begin{eqnarray}\label{reconstruct2_fuv}
f&=&\displaystyle\lim_{\lambda \to \infty}{\lambda\left(\Psi_{0}^{\pm}(\lambda)-\lambda\right)}, \nonumber\\
u&=&-zf-\displaystyle\lim_{\lambda \to \infty}{\lambda\left(\Psi_{1}^{\pm}(\lambda)-x+z\lambda\right)}, \nonumber\\
v_x&=&(1-2t_2 f_y)^{-1}\Big( 2t_2(f_z+u_x f_x)-\nonumber\\
&&\partial_x
\displaystyle\lim_{\lambda \to \infty}{\lambda\left(\Psi_{2}^{\pm}(\lambda)+\lambda^2 t_2-y+2f t_2\right)}\Big),
\end{eqnarray}
are solutions respectively of the nonlinear systems of PDEs (\ref{uvf_system1}) and (\ref{uvf_system2a}),(\ref{uvf_system2b}).

In addition, the first few terms of the $\lambda$ large expansions of the above eigenfunctions of $(\hat L,\hat M_1)$ and
$(\hat L,\hat M_2)$ read, respectively
\beq\label{Psi_lambda_large_1}
\ba{l}
\Psi^{\pm}_0(\lambda)=\lambda+f \lambda^{-1}+\delta^{(2)}_0 \lambda^{-2}+O(\lambda^{-3}), \\
\Psi^{\pm}_1(\lambda)=-z\lambda +x-(u+zf) \lambda^{-1}+\delta^{(2)}_1 \lambda^{-2}+O(\lambda^{-3}), \\
\Psi^{\pm}_2(\lambda)=-t_1\lambda +y-(v+t_1 f) \lambda^{-1}+\delta^{(2)}_2 \lambda^{-2}+O(\lambda^{-3}),
\ea
\eeq
where
\beq
\ba{l}
{\delta^{(2)}_0}_x=-(\partial_z+u_x\partial_x+v_x\partial_y)f, \\
{\delta^{(2)}_1}_x=(\partial_z+u_x\partial_x+v_x\partial_y)(u+zf), \\
{\delta^{(2)}_2}_x=(\partial_{z}+u_x\partial_x+v_x\partial_y)(v+t_1 f),
\ea
\eeq
and
\beq\label{Psi_lambda_large_2}
\ba{l}
\Psi^{\pm}_0(\lambda)=\lambda+f \lambda^{-1}+\delta^{(2)}_0 \lambda^{-2}+O(\lambda^{-3}), \\
\Psi^{\pm}_1(\lambda)=-z\lambda +x-(u+zf) \lambda^{-1}+\delta^{(2)}_1 \lambda^{-2}+O(\lambda^{-3}), \\
\Psi^{\pm}_2(\lambda)=-t_2\lambda^2 +y-2f t_2+\tilde{\delta}^{(1)}_2 \lambda^{-1}+O(\lambda^{-2}),
\ea
\eeq
where
\beq\label{def_red9}
\ba{l}
(\tilde{\delta}^{(1)}_2)_x=2t_2 (f_z+u_x f_x+v_x f_y) -v_x.
\ea
\eeq
The coefficients $\delta^{(2)}_2$ and $\tilde{\delta}^{(1)}_2$ satisfy also the compatible equations
\beq
\ba{l}
(\delta^{(2)}_2)_y=(\partial_{t_1}+u_y\partial_x +v_y\partial_y)(v+t_1 f), \\
(\tilde{\delta}^{(1)}_2)_y=2t_2 (\gamma+u_{y}f_{x}+v_{y}f_{y}) -v_y.
\ea
\eeq

The proof of these results is standard, in the NRH dressing philosophy (see, f. i., \cite{MS_dKP_breaking_08} for details). The NRH dressing for the Dunajski hierarchy was presented in \cite{BDM}, but no connection to the initial data was given.

We remark that the NRH problem (\ref{RH1}) is characterized by the following nonlinear integral equations in the spectral variable $\lambda$:
\beq\label{integral}
\Psi_{j}^{-}(\lambda) = \nu_j + \frac{1}{2 \pi i} \int_{\mathbb{R}} \frac{d \lambda'}{\lambda' -(\lambda - i \varepsilon)} R_{j} (\vec\Psi^{-}(\lambda')),~~j=0,1,2,
\eeq
where $R_j(\vec\zeta)\equiv \vec{\cal R}_j(\vec\zeta)-\zeta_j,~j=0,1,2$, and the reconstruction formulae for $f$ and $u$ in terms of the
RH data read
\beq\label{RH_reconstruction_fu}
\ba{l}
f=-\frac{1}{2\pi i}\int_{\RR}R_0\Big(\vec\Psi^{-}(\lambda)\Big)d\lambda , \\
u=-z f+\frac{1}{2\pi i}\int_{\RR}R_1\Big(\vec\Psi^{-}(\lambda)\Big)d\lambda,
\ea
\eeq
while that for $v$ is respectively
\beq\label{RH_reconstruction_v1}
\ba{l}
v=-t_1 f+\frac{1}{2\pi i}\int_{\RR}R_2\Big(\vec\Psi^{-}(\lambda)\Big)d\lambda,
\ea
\eeq
and
\beq\label{RH_reconstruction_v2}
\ba{l}
v_x=(1-2t_2 f_y)^{-1}\left(  2t_2(f_z+u_x f_x)+
\frac{1}{2\pi i}\partial_x\int_{\RR}R_2\Big(\vec\Psi^{-}(\lambda)\Big)d\lambda \right),
\ea
\eeq
for equations (\ref{uvf_system1}) and (\ref{uvf_system2a}),(\ref{uvf_system2b}).

The reality constraint for the RH data is a nonlinear formula:
\begin{eqnarray} \label{reality}
\vec{\mathcal{R}}\left(\overline{\vec{\mathcal{R}}(\bar{\vec{\xi}})}\right)=\vec{\xi}, ~~\forall \vec{\xi} \in \mathbb{C}^{3}.
\end{eqnarray}
It is possible to show that, if $\vec{\mathcal{R}}$ satisfies (\ref{reality}), then
\beq
\overline{\vec\Psi^{-}(\bar\lambda)}=\vec\Psi^{+}(\lambda),~~~ \lambda\in\RR
\eeq
and $u,v,f\in\RR$ (see, f.i., \cite{MS_dKP_breaking_08} for the proof).

We end our remarks noticing that, in the normalization (\ref{norm2}) corresponding to the second equation of the Dunajski hierarchy, the dependence on the variable $y$ is through the combination $y-2ft_2$. Therefore the eigenfunctions $\vec\Psi^-$ and, through the reconstruction formulae (\ref{RH_reconstruction_fu}),(\ref{RH_reconstruction_v2}), also the solutions $f,u,v$, depend on $y$ through the combination $y-2ft_2$. This is conceptually similar to the case of dKP \cite{MS_dKP_breaking_08}, and provides the spectral mechanism for breaking of smooth and localized solutions of equation (\ref{uvf_system2a}),(\ref{uvf_system2b}) at finite time $t_2$. Such a mechanism is absent in equation (\ref{uvf_system1}), for which a smooth evolution is expected.

\section{Constrained spectral data and reductions}

In this section we show how the eigenfunctions and the spectral data are constrained for the reductions of \S 2.2. We have already investigated in \S 3 and \S 4 the reality reduction. We concentrate now on the divergenceless constraint and on the constraint that the Orlov eigenfunctions are polynomial in $\lambda$.
\vskip 10pt
\noindent
\subsection{The divergenceless constraint}

Consider a basis of three independent eigenfunctions $\psi_j,~j=0,1,2$ of $\hat L$ and construct the $3\times 3$ Jacobian matrix
\beq
M=\left( \frac{\partial(\psi_0,\psi_1,\psi_2)}{\partial(x_0,x_1,x_2)} \right)
\eeq
(of components $M_{ij}=\partial\psi_i /\partial {x_j},~i,j=0,1,2$), and its determinant $J(\vec\psi)=\det M$,
 where $x_0=\lambda,~x_1=x,~x_2=y$. Then the following identity holds true:
\beq
\hat L J=-(u_{xx}+v_{xy})J,
\eeq
implying that, if $\hat L$ is divergenceless, i.e., if $u,v$ satisfy the constraint (\ref{volumeVF}), then $J(\vec\psi)$ is also an eigenfunction of $\hat L$: $\hat L J(\vec\psi)=0$.

In the direct problem philosophy, it is possible to show that, if $u,v$ satisfy the divergenceless constraint (\ref{volumeVF}), then the spectral data
$\vec{\cal S},\vec{\cal K},\vec{\cal R}$ satisfy the following constraint
\beq\label{divergenceless_data}
\det \left(\frac{\partial(D_0,D_1,D_2)}{\partial(\zeta_0,\zeta_1,\zeta_2)} \right)=1.
\eeq
To show it, consider as a basis the Jost solutions $\vec\phi_-$. Then $\hat L J(\vec\phi_-)=0$ and $J(\vec\phi_-)\to 1$ as $z\to-\infty$. Since $1$ is an exact eigenfunction, then $J(\vec\phi_-)=1$. Analogously, $J(\vec\Psi^{\pm})\sim 1$ as $\lambda\sim\infty$; therefore $J(\vec\Psi^{\pm})$ are analytic eigenfunctions going like $1$ at $\lambda\sim\infty$; then $J(\vec\Psi^{\pm})= 1$. Evaluating $J(\vec\phi_-)=1$ at $z=\infty$, we infer the constraint (\ref{divergenceless_data}) for the scattering data $\vec{\cal S}$. In addition, from equations (\ref{relation1}) and (\ref{relation2}), it follows that
\beq
J(\vec\Psi^{\pm})=\det\left(\frac{\partial({\cal K}^{\pm}_0,{\cal K}^{\pm}_1,{\cal K}^{\pm}_2)}{\partial(\zeta_0,\zeta_1,\zeta_2)}  \right)J(\vec\phi_-),
\eeq
implying the constraint (\ref{divergenceless_data}) for the spectral data $\vec{\cal K}$. At last, from the NRH problem (\ref{RH2}), it follows that
\beq\label{linearized_RH}
J(\vec\Psi^{+})=\det\left(\frac{\partial(\cal R_0,\cal R_1,\cal R_2)}{\partial(\zeta_0,\zeta_1,\zeta_2)}  \right)J(\vec\Psi^{-}),
\eeq
implying the constraint (\ref{divergenceless_data}) for the RH data $\vec{\cal R}$.

Viceversa, in the inverse problem, if the RH data satisfy the constraint (\ref{divergenceless_data}), i.e., the NRH problem (\ref{RH1}) is a volume preserving mapping, then $u,v$, reconstructed via (\ref{reconstruct1_fuv}),(\ref{reconstruct2_fuv}), satisfy the divergenceless constraint (\ref{volumeVF}). The proof is standard (see, f.i., \cite{BDM}): if the RH data $\vec{\mathcal{R}}(\vec{\zeta})$ satisfy the volume preserving constraint (\ref{divergenceless_data}),
then, from (\ref{linearized_RH}), it follows that $J(\vec\Psi^{+})=J(\vec\Psi^{-})$, providing the analytic continuation one of the other. Therefore $J(\vec\Psi^{+})$ is entire in $\lambda$, and since $J(\vec\Psi^{+})\to 1$ as $\lambda\to\infty$, it follows that that $J(\vec\Psi^{+})=J(\vec\Psi^{-})=1$. As a consequence of it, the coefficient $-(u_x+v_y)$ of the $O(\lambda^{-1})$ expansion of $J(\vec\Psi^{+})$ must be zero, and the constraint (\ref{volumeVF}) follows. $\Box$ \\
\\
\noindent
\subsection{Polynomial Orlov eigenfunctions}

Let $g(\lambda)$ be an eigenfunction of $\hat L$ characterized by the formal expansion
\begin{eqnarray}
g(\lambda) \sim \sum_{n=-\infty}^{N}g_{n} \lambda^{n},~|\lambda| \gg 1, \ \ N>0.
\end{eqnarray}
Then the coefficients $g_{n}$ satisfy the recursion relations
\begin{eqnarray} \label{recursion}
0&=&g_{N,x}, \nonumber \\
0&=&g_{N-1,x}+(\partial_z+\hat u_x)g_{N},\nonumber\\
0&=&g_{n-1,x}-(n+1)f_{x}g_{n+1}+(\partial_z+\hat u_x)g_{n},~~~n < N.
\end{eqnarray}
This three term recursion implies that, if $g_{-1}=0$, then $g_n=0$ for any $n\le -1$. Consequently $g_{-1}=0$ is an admissible constraint for the dPDEs associated with $\hat L$, and corresponds to the existence of polynomial eigenfunctions $g(\lambda)$.

Let us apply this mechanism to the Orlov-type eigenfunctions $\Psi^{\pm}_1$ and $\Psi^{\pm}_2$ of the first two members of the hierarchy.
The reduction (\ref{red1}) is the $g_{-1}=0$ condition for the eigenfunctions $\Psi^{\pm}_1$ (see (\ref{Psi_lambda_large_1}),(\ref{Psi_lambda_large_2})); therefore it corresponds to the following elementary constraints on the eigenfunctions and data:
\beq
\Psi^{\pm}_1=x-\lambda z~~\Leftrightarrow~~R_1=0
\eeq
for both dynamics (\ref{uvf_system1}) and (\ref{uvf_system2a}),(\ref{uvf_system2b}). The reduction (\ref{red2}) is the $g_{-1}=0$ condition for the eigenfunctions $\Psi^{\pm}_2$ for the dynamics (\ref{uvf_system1}), and it corresponds to
\beq
\Psi^{\pm}_2=y-\lambda t_1~~\Leftrightarrow~~R_2=0.
\eeq
The $g_{-1}=0$ condition for the eigenfunctions $\Psi^{\pm}_2$ for the dynamics (\ref{uvf_system2a}),(\ref{uvf_system2b}) corresponds instead to the differential reduction (\ref{red9}) (see (\ref{Psi_lambda_large_2}),(\ref{def_red9})); it corresponds to the constraints
\beq
\Psi^{\pm}_2=-\lambda^2 t_2+y-2 f t_2~~\Leftrightarrow~~R_2=0.
\eeq
Using the above constraints on the RH data, we can characterize the solutions of the reductions presented in \S 2 in terms of NRH problems.\\

{\it 1. Consider the NRH problem (\ref{RH2}) and let the RH data satisfy the volume preserving
condition
\begin{eqnarray}\label{C_Dunaj}
\det\left(\frac{\partial(\cal R_0(\vec\zeta),\cal R_1(\vec\zeta),\cal R_2(\vec\zeta))}{\partial(\zeta_0,\zeta_1,\zeta_2)}  \right)=1;
\end{eqnarray}
then $f$ and $\theta$, defined by
\beq\label{recons1_ftheta}
\ba{l}
f=-\frac{1}{2\pi i}\int_{\RR}R_0\Big(\Psi^{-}_0(\lambda),\Psi^{-}_1(\lambda),\Psi^{-}_2(\lambda)\Big)d\lambda, \\
\theta_{y}=-z f+\frac{1}{2\pi i}\int_{\RR}R_1\Big(\Psi^{-}_0(\lambda),\Psi^{-}_1(\lambda),\Psi^{-}_2(\lambda)\Big)d\lambda,
\ea
\eeq
solve the Dunajski equation (\ref{Dunaj}) \cite{BDM} and equations (\ref{red6_abg}) for, respectively, the normalizations (\ref{norm1}) and (\ref{norm2}). \\
If, in addition, $R_0=0$, then $f=0$, $\Psi^{\pm}_0=\lambda$, (\ref{RH2}) becomes the $2$-vector NRH problem
\beq
\ba{l}
\Psi^{+}_1(\lambda)={\cal R}_1\Big(\lambda,\Psi^{-}_1(\lambda),\Psi^{-}_2(\lambda)\Big), \ \ \
\Psi^{+}_2(\lambda)={\cal R}_2\Big(\lambda,\Psi^{-}_1(\lambda),\Psi^{-}_2(\lambda)\Big),
\ea
\eeq
the RH data satisfy the constraint
\begin{eqnarray}\label{C_Dunaj}
\det\left(\frac{\partial(\cal R_1(\vec\zeta) ,\cal R_2(\vec\zeta) )}{\partial(\zeta_1,\zeta_2)}  \right)=1,
\end{eqnarray}
and
\beq\label{recons1_theta}
\ba{l}
\theta_{y}=\frac{1}{2\pi i}\int_{\RR}R_1\Big(\lambda,\Psi^{-}_1(\lambda),\Psi^{-}_2(\lambda)\Big)d\lambda
\ea
\eeq
solves the heavenly equation (\ref{heavenly}) \cite{MS_IST_heav_06} and equations (\ref{second_heavenly}) for, respectively, the normalizations (\ref{norm1}) and (\ref{norm2}).
} \\

{\it 2. Let $R_1=0$; then $u=-zf$, $\Psi^{\pm}_1=x-\lambda z$, (\ref{RH2}) becomes the $2$-vector NRH problem
\beq
\ba{l}
\Psi^{+}_0(\lambda)={\cal R}_0\Big(\Psi^{-}_0(\lambda),x-\lambda z,\Psi^{-}_2(\lambda)\Big), \\
\Psi^{+}_2(\lambda)={\cal R}_2\Big(\Psi^{-}_0(\lambda),x-\lambda z,\Psi^{-}_2(\lambda)\Big),
\ea
\eeq
and $f,v$, constructed via
\beq\label{recons2_fv}
\ba{l}
f=-\frac{1}{2\pi i}\int_{\RR}R_0\Big(\Psi^{-}_0(\lambda),x-\lambda z,\Psi^{-}_2(\lambda)\Big)d\lambda, \\
v=-t_1 f+\frac{1}{2\pi i}\int_{\RR}R_2\Big(\Psi^{-}_0(\lambda),x-\lambda z,\Psi^{-}_2(\lambda)\Big)d\lambda.
\ea
\eeq
and via
\begin{eqnarray}\label{recons7_fv}
f&=&-\frac{1}{2\pi i}\int_{\RR}R_0\Big(\Psi^{-}_0(\lambda),x-\lambda z,\Psi^{-}_0(\lambda)\Big)d\lambda, \nonumber\\
v_x&=&(1-2t_2 f_y)^{-1}\left(  2t_2(f_z-z f^2_x)+
\frac{1}{2\pi i}\partial_x\int_{\RR}R_2\Big(\Psi^{-}_0(\lambda),x-\lambda z,\Psi^{-}_2(\lambda)\Big)d\lambda \right)\nonumber\\
&&
\end{eqnarray}
solve equations (\ref{red1_equ}) and (\ref{red7_equ}) for, respectively, the normalizations (\ref{norm1}) and (\ref{norm2}).\\
If, in addition, the RH data satisfy the volume constraint
\begin{eqnarray}
\det\left(\frac{\partial(\cal R_0(\vec\zeta) ,\cal R_2(\vec\zeta) )}{\partial(\zeta_0,\zeta_2)}  \right)=1,
\end{eqnarray}
then (\ref{comb1}) are satisfied and
\begin{eqnarray}\label{recons7_fv}
\tilde\theta_y&=&\frac{1}{2\pi i}\int_{\RR}R_0\Big(\Psi^{-}_0(\lambda),x-\lambda z,\Psi^{-}_2(\lambda)\Big)d\lambda, \nonumber\\
\end{eqnarray}
solves equations (\ref{equ_theta_tilde}) and (\ref{red8_equ}) for, respectively, the normalizations (\ref{norm1}) and (\ref{norm2}).
}

{\it 3. Let $R_2=0$; then $v=-t_1 f$, $\Psi^{\pm}_2=\nu^{(1)}_2(\lambda)$ for the normalization (\ref{norm1}), and $v$ satisfies the constraint (\ref{red9}) and $\Psi^{\pm}_2=\nu^{(2)}_2(\lambda)$ for the normalization (\ref{norm2}), where of course
\beq
\nu^{(1)}_2(\lambda)=y-\lambda t_1, \ \ \nu^{(2)}_2(\lambda)=-\lambda^2 t_2 +y-2f t_2.
\eeq
The NRH problem (\ref{RH2}) becomes the $2$-vector NRH problems
\beq
\ba{l}
\Psi^{+}_0(\lambda)={\cal R}_0\Big(\Psi^{-}_0(\lambda),\Psi^{-}_1(\lambda),\nu^{(n)}_2(\lambda)\Big), \\
\Psi^{+}_1(\lambda)={\cal R}_1\Big(\Psi^{-}_0(\lambda),\Psi^{-}_1(\lambda),\nu^{(n)}_2(\lambda)\Big), n=1,2
\ea
\eeq
for the normalizations (\ref{norm1}) and (\ref{norm2}), and
\beq\label{recons3_fu}
\ba{l}
f=-\frac{1}{2\pi i}\int_{\RR}R_0\Big(\Psi^{-}_0(\lambda),\Psi^{-}_1(\lambda),\nu^{(n)}_2(\lambda)\Big)d\lambda, \\
u=-z f+\frac{1}{2\pi i}\int_{\RR}R_1\Big(\Psi^{-}_0(\lambda),\Psi^{-}_1(\lambda),\nu^{(n)}_2(\lambda)\Big)d\lambda
\ea
\eeq
are solutions of (\ref{red2_equ}) and (\ref{red9_equ}) for respectively $n=1$ and $n=2$.

If $R_1=R_2=0$, then we have also $u=-zf$, $\Psi^{\pm}_1=x-\lambda f$, and (\ref{RH2}) becomes the scalar NRH problems
\beq
\Psi^{+}_0={\cal R}_0(\Psi^{-}_0,x-\lambda z,\nu^{(n)}_2(\lambda)), \ n=1,2
\eeq
for, respectively, the normalizations (\ref{norm1}) and (\ref{norm2}). At last,
\beq\label{recons11_f}
\ba{l}
f=-\frac{1}{2\pi i}\int_{\RR}R_0\Big(\Psi^{-}_0,x-\lambda z,\nu^{(n)}_2(\lambda)\Big)d\lambda, \ n=1,2
\ea
\eeq
solve respectively the scalar dPDE (\ref{equ_f}) and equation (\ref{red10_equ}).

If $R_2=0$ and the RH data satisfy the volume constraint
\begin{eqnarray}
\det\left(\frac{\partial(\cal R_0(\vec\zeta) ,\cal R_1(\vec\zeta) )}{\partial(\zeta_0,\zeta_1)}  \right)=1,
\end{eqnarray}
then $f$ and $\theta$, defined by
\beq\label{recons1_ftheta}
\ba{l}
f=-\frac{1}{2\pi i}\int_{\RR}R_0\Big(\Psi^{-}_0(\lambda),\Psi^{-}_1(\lambda),\nu^{(2)}_2(\lambda)\Big)d\lambda, \\
\theta_{y}=-z f+\frac{1}{2\pi i}\int_{\RR}R_1\Big(\Psi^{-}_0(\lambda),\Psi^{-}_1(\lambda),\nu^{(2)}_2(\lambda)\Big)d\lambda,
\ea
\eeq
solve equations (\ref{red6_equ}).
}

\section{The longtime behaviour of the solutions}

As in the classical IST method, the two inverse problems for dPDEs are expressed in terms of integral equations in the spectral variable $\lambda$, and the
dependent and independent variables of the nonlinear dPDE appear there just as parameters.
Therefore they are convenient tools to construct the longtime behavior of the solutions of the dPDEs, and this
section is devoted to this goal, concentrating on the NRH problem.

We start with the nonlinear integral equations (\ref{integral}), characterizing the NRH problems associated respectively with the dynamics (\ref{uvf_system1}) and (\ref{uvf_system2a}),(\ref{uvf_system2b}), corresponding respectively to the normalizations (\ref{norm1}) and (\ref{norm2}), and we rewrite it in a more convenient form
\beq\label{integral_Phi}
\Phi_{j}(\lambda) = \frac{1}{2 \pi \mathbf{i}} \int_{\mathbb{R}} \frac{d \lambda'}{\lambda' -(\lambda - \mathbf{i}\varepsilon)} R_{j} (\vec\nu + \vec\Phi),~~j=0,1,2
\eeq
in terms of the functions $\Phi_j$, defined by
\beq
\Phi_j(\lambda):=\Psi^-_j(\lambda)-\nu_j(\lambda),~~j=0,1,2.
\eeq
Consider first the case (\ref{norm1}), and
\beq\label{region_1}
t_{1} \gg 1,~~ x=v_1 t_1,~~y=v_2 t_1 +\tilde y,~~z=v_3 t_1,~~ \tilde y,v_1,v_2,v_3=O(1)
\eeq
Then equations (\ref{integral_Phi}) read
\beq\label{integral_Phi_2}
\ba{l}
\Phi_{j}(\lambda) = \frac{1}{2 \pi \mathbf{i}} \int_{\mathbb{R}} \frac{d \lambda'}{\lambda' -(\lambda - \mathbf{i}\varepsilon)} R_{j}
\Big(\lambda'+\Phi_0(\lambda'),-v_3(\lambda'-v_1/v_3)t_1+\Phi_1(\lambda'),\\
-(\lambda'-v_2)t_1 +\tilde y+\Phi_2(\lambda')\Big),~~j=0,1,2
\ea
\eeq
For $t>>1$, the decay of the integral is partially contrasted if $v_2=v_1/v_3$ or, equivalently, on the surface
\beq\label{region_1a}
y-\frac{x}{z}t_1=\tilde y
\eeq
of the $(x,y,z)$ space, on which the second and third argument of $R_j~(j=0,1,2)$ in (\ref{integral_Phi_2}) have the same linear growth in $t_1$. On such a surface, since the main contribution to the integrals occurs when $\lambda'\sim v_2$, it is convenient to make the change of variables $\mu'=(\lambda'-v_2)t_1$, obtaining, for $t_1>>1$:
\beq\label{integral_Phi_3}
\ba{l}
\Phi_{j}(\lambda) \sim \frac{1}{2 \pi \mathbf{i} t_1} \int_{\mathbb{R}} \frac{d \mu'}{v_2+\frac{\mu'}{t_{1}} -(\lambda - \mathbf{i}\varepsilon)} R_{j}
\Big(v_2+\Phi_0(v_2+\frac{\mu'}{t_1}),-v_3\mu'+\Phi_1(v_2+\frac{\mu'}{t_1}\Big), \\ -\mu' +\tilde y+\Phi_2(v_2+\frac{\mu'}{t_1})),~~j=0,1,2.
\ea
\eeq
If $|\lambda -v_2 |>> O(t_1^{-1})$, equation (\ref{integral_Phi_3}) implies that $\Phi_{j}(\lambda)=O(t_1^{-1})$:
\beq\label{integral_Phi_4}
\ba{l}
\Phi_{j}(\lambda) \sim -\frac{1}{2 \pi \mathbf{i} t_1(\lambda-v_2 -\mathbf{i}\varepsilon)} \int_{\mathbb{R}} d \mu R_{j}
\Big(v_2+\Phi_0(v_2+\frac{\mu}{t_1}),-v_3\mu+\Phi_1(v_2+\frac{\mu}{t_1}),\\ -\mu +\tilde y+\Phi_2(v_2+\frac{\mu}{t_1})\Big),~~j=0,1,2,
\ea
\eeq
while, for $\lambda -v_2=\mu/t_1,~\mu=O(1)$, then $\Phi_{j}(\lambda)=O(1)$ and
\beq\label{integral_Phi_5}
\ba{l}
\Phi_j(v_2+\mu/t_1) \sim \frac{1}{2 \pi \mathbf{i}} \int_{\mathbb{R}} \frac{d \mu'}{\mu' -(\mu - \mathbf{i}\varepsilon)} R_{j}
\Big(v_2+\Phi_0(v_2+\frac{\mu'}{t_1}),-v_3\mu'+\Phi_1(v_2+\frac{\mu'}{t_1}),\\ -\mu' +\tilde y+\Phi_2(v_2+\frac{\mu'}{t_1}\Big),~~j=0,1,2.
\ea
\eeq
Equation (\ref{integral_Phi_4}) for $|\lambda |>>1$ and (\ref{reconstruct1_fuv}) finally imply the following result. \\
In the space-time regions (\ref{region_1}),(\ref{region_1a}), the longtime behavior of the solution $(f,u,v)$ of (\ref{uvf_system1}) reads
\beq
\ba{l}
f\sim \frac{1}{t_1}F(\frac{x}{z},\frac{z}{t_1}, y-\frac{x}{z}t_1)+o(t_1^{-1}), \\
u=-\frac{z}{t_1}F(\frac{x}{z},\frac{z}{t_1}, y-\frac{x}{z}t_1)+o(1),\\
v=-F(\frac{x}{z},\frac{z}{t_1}, y-\frac{x}{z}t_1)+o(1),
\ea
\eeq
where
\beq\label{F_def1}
\ba{l}
F(\xi,\zeta,\eta):=-\frac{1}{2 \pi \mathbf{i}}\int_{\mathbb{R}} d \mu R_{0}
(\xi +A_0(\mu;\xi,\zeta,\eta),-\zeta \mu+A_1(\mu;\xi,\zeta,\eta), \\ -\mu +\eta +A_2(\mu;\xi,\zeta,\eta)),
\ea
\eeq
and the fields $A_j(\mu)=\Phi_j(v_2+\mu/t_1),~j=0,1,2$ are defined by the nonlinear integral equations
\beq\label{integral_A_1}
\ba{l}
A_j(\mu;\xi,\zeta,\eta) = \frac{1}{2 \pi \mathbf{i}} \int_{\mathbb{R}} \frac{d \mu'}{\mu' -(\mu - \mathbf{i}\varepsilon)} R_{j}
\Big(\xi+A_0(\mu';\xi,\zeta,\eta),-\zeta \mu'+ \\ A_1(\mu';\xi,\zeta,\eta),-\mu' +\eta+A_2(\mu';\xi,\zeta,\eta)\Big),~~j=0,1,2.
\ea
\eeq
We observe that the longtime behavior of the solution of (\ref{uvf_system1}) has the same structure as the solution of the linearized version of equation (\ref{uvf_system1}), except from the fact that function $F$ is constructed through the solution of a nonlinear system of integral equations. In addition, the solution, concentrated asymptotically on the wave front (\ref{region_1a}), does not break.

The situation drastically changes if we consider the longtime behavior of solutions of (\ref{uvf_system2a}),(\ref{uvf_system2b}) in the space regions
\beq\label{region_2}
t_{2} \gg 1,~~ x=v_1 t_2,~~y=v_2 t_2 +\tilde y,~~z=v_3 t_2,~~ \tilde y,v_1,v_2,v_3=O(1).
\eeq
Indeed, in this case, in the region (\ref{region_2}) the integral equations (\ref{integral_Phi}) become
\beq\label{integral_Phi_2b}
\ba{l}
\Phi_{j}(\lambda) = \frac{1}{2 \pi \mathbf{i}} \int_{\mathbb{R}} \frac{d \lambda'}{\lambda' -(\lambda - \mathbf{i}\varepsilon)} R_{j}
\Big(\lambda'+\Phi_0(\lambda'),-v_3(\lambda'-v_1/v_3)t_2+\Phi_1(\lambda'),\\
-({\lambda'}^2-v_2)t_2 +\tilde y-2ft_2+\Phi_2(\lambda')\Big),~~j=0,1,2
\ea
\eeq
and now the decay of the integral is partially contrasted if $v_2>0$ and $\pm\sqrt{v_2}=v_1/v_3$; i.e., on the surface
\beq\label{region_2b}
\pm\sqrt{\frac{y-\tilde y}{t_2}}=\frac{x}{z},~~\Rightarrow~~     y-\frac{x^2}{z^2}t_2=\tilde y
\eeq
of the $(x,y,z)$ space. On such a surface, let us consider the case $\sqrt{v_{2}}=v_{1}/v_{3}$, it is convenient to make the change of variables $\mu'=(\lambda'-\sqrt{v_2})t_2$ in (\ref{integral_Phi_2b}), obtaining, for $t_2 \gg 1$:
\beq\label{integral_Phi_3b}
\ba{l}
\Phi_{j}(\lambda) \sim \frac{1}{2 \pi \mathbf{i} t_2} \int_{\mathbb{R}} \frac{d \mu'}{\sqrt{v_2}+\frac{\mu'}{t} -(\lambda - \mathbf{i}\varepsilon)} R_{j}
\Big(\sqrt{v_2}+\Phi_0(\sqrt{v_2}+\frac{\mu'}{t_2}),-v_3\mu'+\Phi_1(\sqrt{v_2}+\frac{\mu'}{t_2}), \\ -2\sqrt{v_2}\mu' +\tilde y-2f t_{2}+\Phi_2(\sqrt{v_2}+\frac{\mu'}{t_2})\Big),~~j=0,1,2.
\ea
\eeq
If $|\lambda -\sqrt{v_2} |>> O(t_2^{-1})$, the  $\Phi_j'$ are $O(t_2^{-1})$ and  equation (\ref{integral_Phi_3}) becomes
\beq\label{integral_Phi_4b}
\ba{l}
\Phi_{j}(\lambda) \sim -\frac{1}{2 \pi \mathbf{i} t_2(\lambda-\mathbf{i}\varepsilon)} \int_{\mathbb{R}} d \mu R_{j}
\Big(\sqrt{v_2}+\Phi_0(\sqrt{v_2}+\frac{\mu}{t_2}),-v_3\mu+ \\ \Phi_1(\sqrt{v_2}+\frac{\mu}{t_2}),-2\sqrt{v_2}\mu +\tilde y-2f t_{2}+\Phi_2(\sqrt{v_2}+\frac{\mu}{t_2})\Big),~~j=0,1,2,
\ea
\eeq
while, if  $\lambda -\sqrt{v_2}=\mu/t_2,~\mu=O(1)$, the $\Phi_j's$ are $O(1)$:
\beq\label{integral_Phi_5b}
\ba{l}
\Phi_{j}(\sqrt{v_2}+\frac{\mu}{t_2}) \sim \frac{1}{2 \pi \mathbf{i}} \int_{\mathbb{R}} \frac{d \mu'}{\mu' -(\mu - \mathbf{i}\varepsilon)} R_{j}
\Big(\sqrt{v_2}+\Phi_{0}(\sqrt{v_2}+\frac{\mu'}{t_2}),-v_3\mu'+ \\ \Phi_{1}(\sqrt{v_2}+\frac{\mu'}{t_2}),-2\sqrt{v_2}\mu' +\tilde y-2f t_{2}+\Phi_{2}(\sqrt{v_2}+\frac{\mu'}{t_2})\Big),~~j=0,1,2.
\ea
\eeq
Equation (\ref{integral_Phi_4b}) for $|\lambda |>>1$ and (\ref{reconstruct2_fuv}) finally imply the following result. \\
In the space-time regions (\ref{region_2}),(\ref{region_2b}) the longtime behavior of the solution $(f,u,v)$ of
(\ref{uvf_system2a}),(\ref{uvf_system2b}) reads
\beq\label{fuv_asympt_2}
\ba{l}
f\sim \frac{1}{t_2}F(\frac{x}{z},\frac{z}{t_2}, y-\frac{x^2}{z^2}t_2 -2ft_2)+o(t_2^{-1}), \\
u=-\frac{z}{t_2}F(\frac{x}{z},\frac{z}{t_2}, y-\frac{x^2}{z^2}t_2 -2ft_2)+o(1),\\
v_x=\frac{2t_2(f_z+u_xf_x)}{1-2t_2 f_y}+o(1),
\ea
\eeq
where $F$ is defined by
\beq\label{def_F2}
\ba{l}
F(\xi,\zeta,\eta)=-\frac{1}{2 \pi \mathbf{i}}\int_{\mathbb{R}} d \mu R_{0}\Big(\xi+A_0(\mu;\xi,\zeta,\eta), -\zeta\mu+A_1(\mu;\xi,\zeta,\eta),\\ -2\xi \mu+\eta+A_2(\mu;\xi,\zeta,\eta) \Big),
\ea
\eeq
and the fields $A_j,~j=0,1,2$ are defined by the nonlinear integral equations
\beq\label{integral_A_2}
\ba{l}
A_{j}(\mu;\xi,\zeta,\eta) = \frac{1}{2 \pi \mathbf{i}} \int_{\mathbb{R}} \frac{d \mu'}{\mu' -(\mu - \mathbf{i}\varepsilon)} R_{j}
\Big(\xi+A_{0}(\mu';\xi,\zeta,\eta),-\zeta\mu'+ \\ A_{1}(\mu';\xi,\zeta,\eta),-2\xi\mu' +\eta +A_{2}(\mu';\xi,\zeta,\eta)\Big),~~j=0,1,2.
\ea
\eeq
By considering the case $\sqrt{v_{2}}=-v_{1}/v_{3}$, we obtain the same result as above.
Equations (\ref{fuv_asympt_2}) show that, even in the longtime regime, localized solutions of equations (\ref{uvf_system2a}) and (\ref{uvf_system2b}) break on the wave front surface (\ref{region_2b}) in a way similar to the dKP breaking \cite{MS_dKP_breaking_08}.

\section{Concluding remarks}

In this paper we have applied the formal Inverse Spectral Transform for integrable dispersionless PDEs to the so-called Dunajski hierarchy. We concentrated, in particular, i) on the system of PDEs characterizing a general anti-self-dual conformal structure in neutral signature, ii) on its first commuting flow, and iii) on some of their basic and novel reductions. We have formally solved their Cauchy problem and used it to construct the longtime behavior of solutions, showing, in particular, that unlike the case of soliton PDEs, different dispersionless PDEs belonging to the same hierarchy of commuting flows evolve in time in very different ways, exhibiting either a smooth dynamics or a gradient catastrophe at finite time.\\
In a subsequent paper, we plan  i) to investigate the analytical aspects of such a wave breaking, taking place, in the longtime regime, on the 3-dimensional surface (\ref{region_2b}); ii) to construct, following \cite{MS_solv_11}, distinguished classes of implicit solutions of the Dunajski hierarchy and of its reductions, concentrating also on the class of reductions introduced in \cite{Bogdanov2}.

\bigskip
\noindent
\textbf{Acknowledgements:} Ge Yi has been supported, in the first part of this research, by the China Scholarship Council program and, in the second part of it, by the Scientific Research Funds of Huaqiao University (14BS308).  Ge Yi and P. M. Santini would like to acknowledge the warm hospitality of the Centro Internacional de Ciencias, Cuernavaca, Mexico, where the last part of this paper was written.


\begin{thebibliography}{99}

\bibitem{AC} M. J. Ablowitz and P. A. Clarkson, \textit{Solitons, nonlinear evolution equations and Inverse Scattering}, London Math. Society Lecture Note 1991.

\bibitem{AS} M. J. Ablowitz and H. Segur \textit{Solitons and the Inverse Scattering Transform} SIAM 1981.


\bibitem{Bogdanov1} L. V. Bogdanov ``On a class of reductions of Manakov-Santini hierarchy connected with the interpolating system'', J. Phys. A: Math. Theor. {\bf 43},
(2010), 115206 (11pp).

\bibitem{BDM} L. V. Bogdanov, V. S. Dryuma and S. V. Manakov: ``Dunajski generalization of the second heavenly equation: dressing method and the hierarchy'', J. Phys. A: Math. Theor. 40, (2007), 14383-14393.

\bibitem{Bogdanov2} L. V. Bogdanov ``Interpolating differential reductions of multidimensional integrable hierarchies'', Theor. Math. Phys., {\bf 167} (3), (2011), 705-713.

\bibitem{BK} L. V. Bogdanov and B. G. Konopelchenko, ``On the $\bar\partial$-dressing method applicable to heavenly equation''; Phys. Lett. A {\bf 345}, (2005) 137-143.

\bibitem{BK2} L. V. Bogdanov and B. G. Konopelchenko, ``Grassmannians Gr(N-1,N+1), closed differential N-1 forms and N-dimensional integrable systems'', J. Phys. A: Math. Theor. 46 (2013) 085201.

\bibitem{BF} C. Boyer and J. D. Finley, ``Killing vectors in self-dual, Euclidean Einstein spaces'',
J. Math. Phys. {\bf 23} (1982), 1126-1128.



\bibitem{CD} F. Calogero and A. Degasperis {\it Spectral Transform and Solitons} North-Holland Publishing Company, 1982.

\bibitem{Dunaj} M. Dunajski, ``Anti-self-dual four manifolds with a parallel real spinor'', Proc. R. Soc. A {\bf 458}, (2002), 1205-1222.

\bibitem{Duna} M. Dunajski, ``A class of Einstein-Weyl spaces associated to an integrable system of hydrodinamic type'', J. Geom. Phys. {\bf 51}  (2004), 126-137.

\bibitem{DuFerK} M. Dunajski, E. V. Ferapontov and B. Kruglikov, ``On the Einstein-Weyl and conformal self-duality equations''; arXiv:1406.0018v2.

\bibitem{DuMa1} M. Dunajski and L. J. Mason, ``Hyper-K\"ahler hierachies and their twistor theory'', Comm. Math. Phys. {\bf 213}, (2000), 641-672.

\bibitem{DuMa2} M. Dunajski and L. J. Mason, ``Twistor theory of hyper-K\"ahler metrics with hidden symmetries'', J. Math. Phys., {\bf 44},(2003), 3430-3454.

\bibitem{DMT} M. Dunajski, L. J. Mason and K. P. Tod, ``Einstein-Weyl geometry, the dKP equation and twistor theory'', J. Geom. Phys. {\bf 37} (2001), 63-93.

\bibitem{DT} M. Dunajski and K. P. Tod, ``Einstein-Weyl spaces and dispersionless Kadomtsev-Petviashvili equation from Painlev\'e I and II''; arXiv:nlin.SI/0204043.

\bibitem{Ferapontov} E. V. Ferapontov and K. R. Khusnutdinova: ``On integrability of (2+1)-dimensional quasilinear systems'', Comm. Math. Phys. {\bf 248} (2004) 187-206.

\bibitem{FP} J. D. Finley and J. F. Plebanski: ``The classification of all ${\cal K}$ spaces admitting
a Killing vector'', J. Math. Phys. {\bf 20}, 1938 (1979).


\bibitem{Gakhov} F. D. Gakhov, \textit{Boundary value problems}, Dover, New York (1990).


\bibitem{GGKM} C. S. Gardner, J. M. Greene, M. D. Kruskal and R. M. Miura, ``Method for Solving the Korteweg-deVries Equation'', Phys. Rev. Lett.,
{\bf 19}, (1967), 1095-1097.

\bibitem{GD} J. D. Gegenberg and A. Das, ``Stationary Riemaniann space-times with self-dual curvature'',
Gen. Rel. Grav. {\bf 16} (1984), 817-829.


\bibitem{G-M-MA} F. Guil, M. Manas and L. Martinez Alonso, ``On twistor solutions of the dKP equation'', J. Phys. A:Math. Gen. {\bf 36} (2003) 6457-6472.

\bibitem{GS} P. G. Grinevich and P. M. Santini: ``Holomorphic eigenfunctions of the vector field
associated with the dispersionless Kadomtsev - Petviashvili equation'',  J. Differential Equations,
{\bf 255}, 7, 1469-1491 (2013).

\bibitem{GSW} P. G. Grinevich, P. M. Santini and D. Wu: ``The Cauchy problem for the Pavlov equation'', Preprint arXiv:1310.5834v2.

\bibitem{H} N. J. Hitchin, ``Complex manifolds and Einstein's equations'', in {\it Twistor Geometry and
Nonlinear Systems}, H. D. Doebner and T. Weber (eds), Lecture Notes in Mathematics, vol. 970 (Springer-Verlag
1982).

\bibitem{J} P. E. Jones and K. P. Tod, ``Minitwistor spaces and Einstein-Weyl spaces'', Class. Quantum Grav.
{\bf 2} (1985), 565-577.


\bibitem{KdV} D. J. Korteweg and G. de Vries, ``On the change of form of long waves advancing in a rectangular canal, and on a new type of long stationary waves'', Phil. Mag. \textbf{39} (1895), 422--443.

\bibitem{KG} Y. Kodama and J. Gibbons, ``Integrability of the dispersionless KP hierarchy'',
Proc. 4th Workshop on Nonlinear and Turbulent Processes in Physics, World Scientific, Singapore 1990.

\bibitem{K-MA-R} B. Konopelchenko, L. Martinez Alonso and O. Ragnisco, ``The $\bar\partial$-approach for the dispersionless KP hierarchy'', J.Phys. A: Math. Gen. {\bf 34}
(2001) 10209-10217.


\bibitem{KP} B. B. Kadomtsev and V. I. Petviashvili,  ``On the stability of solitary waves in weakly dispersive media'', Sov. Phys. Dokl., {\bf 15}, (1970), 539-541.

\bibitem{KM1} B. Konopelchenko and F. Magri, ``Coisotropic deformations of associative algebras and dispersionless integrable hierarchies'', Comm. Math. Phys. {\bf 274}, (2007), 627-658.

\bibitem{KM2} B. Konopelchenko and F. Magri, ``Dispersionless integrable equations as coisotropic deformations. Extensions and reductions'', Theoretical and Mathematical Physics, {\bf 151}(3), (2007), 803-819.


\bibitem{Kri0} I. M. Krichever, ``Method of averaging for two-dimensional ``integrable'' equations'', Functional Analysis and Its Applications, {\bf 22} (1988), 200-213.

\bibitem{Kri1} I. M. Krichever, ``The dispersionless Lax equations and topological minimal models'', Comm. Math. Phys. {\bf 143} (1992), no. 2, 415-429.

 \bibitem{Kri2} I. M. Krichever, ``The $\tau$-function of the universal Witham hierarchy, matrix models and topological field theories'', Comm. Pure Appl. Math. {\bf 47}, 437-475 (1994).

\bibitem{KMZ} I. Krichever, A. Marshakov and A. Zabrodin, ``Integrable structure of the Dirichlet
boundary problem in multiply-connected domains''.  Comm. Math. Phys.  259  (2005),  no. 1, 1-44.


\bibitem{LBW} S.-Y. Lee, E. Bettelheim, P. Wiegmann, ``Bubble break-off in Hele-Shaw flows-singularities
and integrable structures''  Phys. D  219  (2006),  no. 1, 22-34.

\bibitem{MAM} L. Martinez Alonso and E. Medina: ``Regularisation of Hele-Shaw flows, multiscaling expansions and
the Painlev\'e I equation'', arXiv:0710.3731.


\bibitem{MA-S} L. Martinez Alonso and A. B. Shabat, ``Towards a theory of differential constraints of a hydrodynamic hierarchy'' J. Nonlinear Math. Phys 10(2) (2003) 229.


\bibitem{MS_shiftedRH_05} S. V. Manakov and P. M. Santini 2005 Inverse Scattering Problem for Vector Fields and the Heavenly Equation {\it Preprint arXiv:nlin/0512043}.

\bibitem{MS_IST_heav_06} S. V. Manakov and P. M. Santini: ``Inverse scattering problem for vector fields and the Cauchy problem for the heavenly equation'', Physics Letters A {\bf 359} (2006) 613-619. http://arXiv:nlin.SI/0604017.

\bibitem{MS_dKP_IST_06} S. V. Manakov and P. M. Santini: ``The Cauchy problem on the plane for the
dispersionless Kadomtsev-Petviashvili equation''; JETP Letters, {\bf 83}, No 10, 462-466 (2006).
http://arXiv:nlin.SI/0604016.

\bibitem{MS_Pavlov_IST_07} S. V. Manakov and P. M. Santini: ``A hierarchy of integrable PDEs in
$2+1$ dimensions associated with $1$ - dimensional vector fields''; Theor. Math. Phys. {\bf 152}(1), 1004-1011 (2007).

\bibitem{MS_dKP_breaking_08} S. V. Manakov and P. M. Santini: ``On the solutions of the dKP equation: the nonlinear
Riemann-Hilbert problem, longtime behaviour, implicit solutions and wave breaking''; J. Phys. A: Math.
Theor. {\bf 41} (2008) 055204 (23pp).

\bibitem{MS_heav_pavlov_09} S. V. Manakov and P. M. Santini: ``On the solutions of the second heavenly and Pavlov
equations'', J. Phys. A: Math. Theor. {\bf 42} (2009) 404013 (11pp). doi: 10.1088/1751-8113/42/40/404013.
ArXiv:0812.3323.

\bibitem{MS_d2DT_09} S. V. Manakov and P. M. Santini, ``The dispersionless 2D Toda equation: dressing, Cauchy problem, longtime behaviour, implicit solutions and wave breaking'', J. Phys. A: Math. Theor. 42, 095203, 16 pages (2009).

\bibitem{MS_solv_11} S. V. Manakov and P. M. Santini, Solvable vector nonlinear Riemann problems, exact implicit solutions of dispersionless PDEs and wave breaking,   J. Phys. A: Math. Theor. 44 345203 (2011).

\bibitem{MS_dKPn_11} S. V. Manakov and P. M. Santini: ``On the dispersionless Kadomtsev-Petviashvili equation in n+1 dimensions: exact solutions, the Cauchy problem for small initial data and wave breaking'', J. Phys. A: Math. Theor. 44 (2011) 405203 (15pp), doi:10.1088/1751-8113/44/40/405203. arXiv:1001.2134.

\bibitem{MS_finite_t_12} S. V. Manakov and P. M. Santini: ``Wave breaking in solutions of the dispersionless
Kadomtsev-Petviashvili equation at finite time'', Theor. Math. Phys. {\bf 172}(2) 1118-1126 (2012).

\bibitem{MS_PMNP13_14} S. V. Manakov and P. M. Santini: ``Integrable dispersionless PDEs arising as commutation
condition of pairs of vector fields'' Proceedings of the conference PMNP 2013. J. Phys.: Conf. Ser.
{\bf 482} (2014) 012029. doi:10.1088/1742-6596/482/1/012029. http://iopscience.iop.org/1742-6596/482/1/012029.
arXiv:1312.2740v1.

\bibitem{MWZ} M. Mineev-Weinstein, P. Wigmann and A. Zabrodin, ``Integrable structure of interface dynamics'',
Phys. Rev. Lett. {\bf 84}, 5106 (2000).


\bibitem{NNS} F. Neyzi, Y. Nutku and M. B. Sheftel, ``A multi-hamiltonian structure of the Plebanski's second heavenly equation''; J. Phys. A: Math. Gen. {\bf 38} (2005),
8473-8485.


\bibitem{Pavlov} M. V. Pavlov: ``Integrable hydrodynamic chains'', J. Math. Phys. {\bf 44} (2003) 4134-4156.

\bibitem{Penrose1976} R. Penrose, Nonlinear gravitons and curved twistor theory, Gen. Rel. Grav 7, 31-52 (1976).

\bibitem{Plebanski} F. Plebanski, Some solutions of complex Einstein equations, J. Math. Phys. 16, 2395-2402 (1975).


\bibitem{Taka1} K. Takasaki, ``Area preserving diffeomorphisms and nonlinear integrable systems'', 1991, in ``Turku 1991, Proceedings, Topological and geometrical methods in field theory 383''.

\bibitem{TT1} K. Takasaki and T. Takebe, ``SDiff(2) Toda equation -- hierarchy, tau function and symmetries'', Lett. Math. Phys. {\bf 23}, (1991), 205-214.

\bibitem{TT2} K. Takasaki and T. Takebe, ``SDiff(2) KP hierarchy'', A. Tsuchiya, T. Eguchi and T. Miwa (eds.), {\it Infinite Analysis\/}, Adv. Ser. Math. Phys. 16 (World Scientific, Singapore, 1992), part B, 889-922.

\bibitem{TT3} K. Takasaki and T. Takebe, ``Integrable hierarchies and dispersionless limit'', Rev. Math. Phys. {\bf 7}, (1995), 743.
\bibitem{Timman1962} R. Timman,  Unsteady motion in transonic flow, \textit{Symposium Transsonicum}, Aachen 1962 ed K. Oswatitsch
(Berlin: Springer), 394-401 (1962).

\bibitem{WZ} P. Wigmann and A. Zabrodin, ``Conformal maps and integrable hierarchies'',
Comm. Math. Phys. {\bf 213}, 523-538 (2000).


\bibitem{Zakharov} V. E. Zakharov, ``Dispersionless limit of integrable systems in 2+1 dimensions'', in {\it Singular Limits of Dispersive Waves}, edited by N.M.Ercolani et al., Plenum Press, New York, 1994.

\bibitem{Ward} R. S. Ward, ''Einstein-Weyl spaces and SU($\infty$) Toda fields'', Class. Quantum Grav.
{\bf 7} (1990) L95-L98.


\bibitem{ZMNP} V. E. Zakharov, S. V. Manakov, S. P. Novikov and L. P. Pitaevsky, \textit{Theory of solitons}, 1984, Plenum Press, New York.

\bibitem{ZS} V. E. Zakharov and A. B. Shabat, ``Integration of nonlinear equations of mathematical physics by the method of inverse scattering. II'',
Functional Anal. Appl. {\bf 13}, (1979), 166-174.

\bibitem{Zak} V. E. Zakharov: ``Integrable systems in multidimensional spaces'', Lecture Notes in Physics, Springer-Verlag, Berlin {\bf 153} (1982), 190-216.

\bibitem{Zobolotskaya1969} E. A. Zobolotskaya and R. V. Kokhlov, Quasi-plane waves in the nonlinear acoustics of confined beams,
Sov. Phys. Acoust. 15, 35-40 (1969).

















\end{thebibliography}
\end{document}